\documentclass[anonymous,3p,times,authoryear]{elsarticle}
\usepackage{amssymb}
\usepackage{amsmath}
\usepackage{amsthm}
\usepackage{url}
\usepackage{hyperref}

\journal{Theoretical Computer Science}

\begin{document}

\begin{frontmatter}
\title{Optimal Extended Formulations from Optimal Dynamic Programming Algorithms} 

\author[label1,label2]{Mateus de Oliveira Oliveira}
\author[label1]{Wim Van den Broeck}

\affiliation[label1]{
    organization={Informatics Institute, University of Bergen},
    addressline={Thormøhlens Gate 55},
    city={Bergen},
    postcode={5006},
    country={Norway}
}
\affiliation[label2]{
    organization={Department of Computer and Systems Sciences (DSV), Stockholm University},
    addressline={Borgarfjordsgatan 12},
    city={Kista},
    postcode={164 55},
    country={Sweden}
}

\begin{abstract}
Vertex Subset Problems (VSPs) are a class of combinatorial optimization problems on graphs where the goal is to find a subset of vertices satisfying a predefined condition. 
Two prominent approaches for solving VSPs are {\em dynamic programming} over tree-like structures, such as tree decompositions or clique decompositions, and {\em linear programming}. In this work, we establish a sharp connection between both approaches by showing that if a vertex-subset problem $\Pi$ admits a solution-preserving dynamic programming algorithm that produces tables of size at most $\alpha(k,n)$ when processing a tree decomposition of width at most $k$ of an $n$-vertex graph $G$, then the polytope $P_{\Pi}(G)$ defined as the convex-hull of solutions of $\Pi$ in $G$ has extension complexity at most $O(\alpha(k,n)\cdot n)$. Additionally, this upper bound is optimal under the exponential time hypothesis (ETH). 

On the one hand, our results imply that ETH-optimal solution-preserving dynamic programming algorithms for combinatorial problems yield optimal-size parameterized extended formulations for the solution polytopes associated with instances of these problems. On the other hand, unconditional lower bounds obtained in the realm of the theory of extended formulations yield unconditional lower bounds on the table complexity of solution-preserving dynamic programming algorithms.
\end{abstract}

\begin{keyword}
Extended Formulations \sep Parameterized Complexity \sep Dynamic Programming Cores
\end{keyword}

\end{frontmatter}


\newcommand{\polytopeP}{P}
\newcommand{\polytopeQ}{Q}
\newcommand{\N}{\mathbb{N}}
\newcommand{\Nplus}{\mathbb{N}_{+}}
\newcommand{\excomfunction}[1]{xc(#1) }
\newcommand{\epoints}[1]{\texttt{endpts}(#1)}
\newcommand{\languageL}{L}
\newcommand{\languageLSlice}[1]{L(#1)}
\newcommand{\SlicePolytope}[1]{\mathrm{P}(#1)}
\newcommand{\polytopex}[1]{\mathrm{P}_{ex}(#1)}
\newcommand{\PolytopeFamily}[1]{\mathcal{P}(#1)}
\newcommand{\conv}{\texttt{conv}}
\newcommand{\setS}{S}
\newcommand{\treeT}{T}
\newcommand{\nodeset}{N}
\newcommand{\inc}{Inc}
\newcommand{\vertexset}{V}
\newcommand{\edgeset}{E}
\newcommand{\increl}{\rho}
\newcommand{\edgee}{e}
\newcommand{\vertexu}{u}
\newcommand{\vertexv}{v}
\newcommand{\graphG}{G}
\newcommand{\graphH}{H}
\newcommand{\graphs}{\textsc{Graphs }}
\newcommand{\isoclosure}[1]{\texttt{ISO}(#1)}
\newcommand{\graphproperty}{\mathbb{P}}
\newcommand{\incomming}{In}
\newcommand{\invariant}{\mathcal{I}}
\newcommand{\arityfunction}{\mathfrak{r}}
\newcommand{\nodep}{p}
\newcommand{\nodeu}{u}
\newcommand{\nodev}{v}
\newcommand{\rootr}{r}
\newcommand{\arca}{a}
\newcommand{\addressmap}{\gamma}
\newcommand{\descendants}[1]{\texttt{desc}(#1)}
\newcommand{\ancestors}[1]{\texttt{anc}(#1)}
\newcommand{\term}{\tau}
\newcommand{\nodes}[1]{N(#1)}
\newcommand{\arcs}[1]{\texttt{Arcs}(#1)}
\newcommand{\symbola}{a}
\newcommand{\symbolb}{b}
\newcommand{\arityr}{\mathrm{r}}
\newcommand{\height}{h}
\newcommand{\stateset}{Q}
\newcommand{\transitionset}{\Delta}
\newcommand{\child}[1]{\texttt{children}(#1)}
\newcommand{\parent}[1]{\texttt{par}(#1)}
\newcommand{\Terms}[1]{\texttt{Terms}(#1)}
\newcommand{\automaton}{\mathcal{A}}
\newcommand{\stateq}{q}
\newcommand{\finalstateset}{F}
\newcommand{\trace}{\rho}
\newcommand{\tracehat}{\hat{\rho}}
\newcommand{\languagemap}{\mathcal{L}}
\newcommand{\graphingfunction}{\mathcal{G}}
\newcommand{\leaf}{\texttt{Leaf}}
\newcommand{\introvertex}[1]{\texttt{IntroVertex}\{#1\}}
\newcommand{\forgetvertex}[1]{\texttt{ForgetVertex}\{#1\}}
\newcommand{\introedge}[2]{\texttt{IntroEdge}\{#1,#2\}}
\newcommand{\join}{\texttt{Join}}
\newcommand{\activelabels}{\mathfrak{b}}
\newcommand{\validterms}{\texttt{ITD}_k}
\newcommand{\labelfunctionset}{\mathcal{F}_k}
\newcommand{\bigcupk}{\bigcup_{k \in \N}\!}
\newcommand{\arcset}{\mathcal{F}}
\newcommand{\witnessbag}{\mathcal{W}}
\newcommand{\final}{\mathrm{Final}}
\newcommand{\clean}{\texttt{Clean}}
\newcommand{\inv}{\texttt{Inv}}
\newcommand{\finitesubsets}[1]{\mathcal{P}_{\texttt{fin}}(#1)}
\newcommand{\dpcore}{\texttt{D}}
\newcommand{\rootfunction}[1]{r(#1)}
\newcommand{\statevarset}{\mathcal{Y}}
\newcommand{\transvarset}{\mathcal{Z}}
\newcommand{\mainvarsinst}{X}
\newcommand{\statevarsinst}{Y}
\newcommand{\transvarsinst}{Z}
\newcommand{\statevar}{y}
\newcommand{\transvar}{z}
\newcommand{\consequent}[1]{\texttt{cnq}(#1)}
\newcommand{\antecedent}[1]{\texttt{ant}(#1)}
\newcommand{\mip}{\mathfrak{L}}
\newcommand{\symbolfunction}[1]{\texttt{symb}(#1)}
\newcommand{\mainvar}{x}
\newcommand{\varset}{\mathcal{X}}
\newcommand{\R}{\mathbb{R}}
\newcommand{\mainvarset}{\mathcal{X}}
\newcommand{\variables}{\mathrm{var}}
\newcommand{\ancvars}{\mathcal{V}}
\newcommand{\taubar}{\Bar{\tau}}
\newcommand{\htau}{\hat{\term}}
\newcommand{\vectorv}{\Bar{v}}
\newcommand{\realvec}{v}
\newcommand{\treedec}{\mathcal{T}}
\newcommand{\lpsoly}{y}
\newcommand{\lpsolx}{x}
\newcommand{\tree}{T}
\newcommand{\bagB}{B}
\newcommand{\nodet}{t}
\newcommand{\treewidth}[1]{tw(#1)}
\newcommand{\tw}{tw}
\newcommand{\leafl}{l}
\newcommand{\colorsq}{q}
\newcommand{\witness}{\mathrm{w}}
\newcommand{\colorc}{c}
\newcommand{\intk}{k}
\newcommand{\functionf}{f}
\newcommand{\functiong}{g}
\newcommand{\functionh}{h}
\newcommand{\cliquenum}{w}
\newcommand{\cliquewidthvalue}{k}
\newcommand{\cw}{cw}
\newcommand{\exprtree}{\mathrm{T}}
\newcommand{\labelsubset}{I}
\newcommand{\labelset}{L}
\newcommand{\bigO}[1]{\mathcal{O}(#1)}
\newcommand{\bigOstar}[1]{\mathcal{O}^*(#1)}
\newcommand{\smallo}[1]{\mathrm{o}(#1)}
\newcommand{\problem}{\Pi}
\newcommand{\solutionset}{\mathrm{Sol}}
\newcommand{\dynamization}{\Gamma}
\newcommand{\algorithmcore}{\mathfrak{A}}
\newcommand{\witnesstree}{W}
\newcommand{\membershipfunction}{\mu}
\newcommand{\treewidthvalue}{k}
\newcommand{\characteristictree}{\chi}
\newcommand{\allcharacteristictrees}{\Phi}
\newcommand{\bagfunction}{B}
\newcommand{\edgefunction}{\xi}
\newcommand{\powerset}[1]{\mathcal{P}(#1)}
\newcommand{\localcomplexity}{\alpha}
\newcommand{\vertexnodemap}{\nu}
\newcommand{\edgenodemap}{\varepsilon}
\newcommand{\extractedvertexset}{\mathbf{X}}
\newcommand{\extractededgeset}{\mathbf{Y}}
\newcommand{\solsetX}{X}
\newcommand{\solsetY}{Y}
\newcommand{\profile}{\mathfrak{p}}
\newcommand{\profileentry}{p}
\newcommand{\solvertexx}{x}
\newcommand{\solvertexy}{y}
\newcommand{\projection}{\Pi}
\newcommand{\convexhull}{\mathrm{conv}}
\newcommand{\traces}{\mathrm{Traces}}
\newcommand{\leafcore}{\mathrm{Leaf}}
\newcommand{\introvertexcore}{\mathrm{IntroVertex}}  
\newcommand{\introedgecore}{\mathrm{IntroEdge}}
\newcommand{\forgetvertexcore}{\mathrm{ForgetVertex}} 
\newcommand{\joincore}{\mathrm{Join}}
\newcommand{\leafnodecore}{\mathrm{Leaf}}  
\newcommand{\connectnodecore}{\mathrm{Connect}}  
\newcommand{\relabelnodecore}{\mathrm{Relabel}}  
\newcommand{\joinnodecore}{\mathrm{Join}}  
\newcommand{\varx}{x}
\newcommand{\vectormu}{\mu}
\newcommand{\width}{\mathrm{w}}

\theoremstyle{plain}
\newtheorem{theorem}{Theorem}[section]      
\newtheorem{lemma}[theorem]{Lemma}
\newtheorem{proposition}[theorem]{Proposition}
\newtheorem{corollary}[theorem]{Corollary}

\theoremstyle{definition}
\newtheorem{definition}[theorem]{Definition}
\newtheorem{observation}[theorem]{Observation}

\renewcommand{\thempfootnote}{\arabic{mpfootnote}}

\section{Introduction}\label{sec:Introduction}
Computational problems arising in a wide variety of sub-fields of artificial intelligence can be formalized as vertex-subset problems, such as {\sc Independent Set}, {\sc Dominating Set},  {\sc Vertex Cover}, etc. Two prominent algorithmic approaches to attack these problems are {\em dynamic programming} over tree-like structures, and linear programming. While the former approach is suitable to solve vertex subset problems on  graphs of small width, such as treewidth \citep{marx2007can} and cliquewidth \citep{courcelle2000linear}, the latter approach yields practical algorithms whenever the space of feasible solutions can be defined using a system of linear inequalities of moderate size \citep{HU2019,Aprile2017}. 

Vertex subset problems that are NP-hard do not admit efficient linear programming formulations assuming 
${P}\neq {NP}$, 
and in some cases it can be even shown unconditionally that formulations for certain problems require exponential size \citep{yannakakis1988expressing,FioriniMassarPokuttaTiwaryWolf2015,DeSimone1990,Watson2018}. Nevertheless, recently there has been growing interest in the study of linear formulations of combinatorial polytopes parameterized by structural parameters of the input graph, such as treewidth and cliquewidth \citep{Seif2019,ABOULKER2019,Buchanan2015,Kolman2020}.
In this work, we contribute with this line of research by establishing a sharp connection between dynamic programming algorithms operating on tree decompositions and linear programming theory.

On the one hand, our results allow us to show that efficient solution-preserving dynamic programming algorithms parameterized by treewidth can be used to define small linear programming formulations. On the other hand, unconditional lower bounds obtained in the realm of the theory of extended formulations can be used to provide unconditional lower bounds on the table-complexity of solution-preserving dynamic programming algorithms expressible in the dynamic programming core model (DP-core model)\citep{baste2022diversity,DBLP:conf/aaai/OliveiraV23}. 
\paragraph{Related recent work}
After the publication of the conference version of this paper, Kluk and Nederlof \citep{kluk2025lowerboundspuredynamic} obtained unconditional lower bounds for "pure" dynamic programming on graphs of bounded pathwidth. In particular, they show that any such dynamic programming algorithm for {\sc Hamiltonian Cycle} requires table size $2^{\Omega(k \log \log k)}$, where $k$ is the pathwidth. Their approach is based on modeling dynamic programming algorithms by tropical circuits and deriving lower bounds via communication complexity through compatibility matrices. This provides a complementary route to unconditional lower bounds on dynamic programming.

\subsection{Our Results}
Our main result (Theorem \ref{theorem:DPToExtendedFormulation}) states that if a vertex subset problem $\problem$ can be solved by a solution-preserving DP-core of table complexity $\localcomplexity(\treewidthvalue,n)$ when processing a width-$\treewidthvalue$ tree decomposition of an $n$-vertex graph $\graphG$, then 
the solution polytope of $\problem$ on $\graphG$ has extension complexity $O(\localcomplexity(\treewidthvalue,n)\cdot n)$.
Intuitively, problems that can be solved by solution-preserving DP-cores with small table complexity yield solution polytopes of small extension complexity. Conversely, lower bounds on the extension complexity of solution polytopes of a given problem $\problem$ yield lower bounds on the table complexity of solution-preserving DP-algorithms solving $\problem$. 

\subsection{Proof Techniques}
We formalize the notion of a dynamic programming algorithm parameterized by treewidth using an extension of the DP-core model introduced in \citep{baste2022diversity} and further developed in \citep{DBLP:conf/aaai/OliveiraV23}.
The main conceptual addition to these models is an axiomatic definition of the notion of a solution-preserving dynamic programming algorithm. Intuitively, this notion formalizes dynamic programming algorithms satisfying two properties: first, any subset of
vertices obtained by backtracking is a solution for the problem in question. Second, every solution should be retrievable by backtracking. It is worth noting that in general, dynamic programming algorithms may not be solution preserving. For instance, if one is simply interested in determining whether a solution exists then some solutions may be discarded during the dynamic programming process. 

The bridge between dynamic programming and linear programming is made within the context of tree automata theory. More specifically, we define the notion of a $\treeT$-shaped tree automaton, where $T$ is a tree. Intuitively, these are automata accepting sets of terms over a given alphabet $\Sigma$, all of which have the same underlying tree-structure. The state set of such an automaton $\automaton$ is partitioned into a collection of cells, one cell $\stateset_{\nodeu}$ per node $\nodeu$ of $\treeT$, and we define the width of $\automaton$ as the size of the largest cell. We then show that if $\automaton$ is a $T$-shaped tree automaton of width $\width(\automaton)$, the polytope $\SlicePolytope{\automaton}$ whose vertices correspond to terms accepted by $\automaton$ has extension complexity $\bigO{|\treeT|\cdot(\width(\automaton) + |\Sigma|)}$ (Corollary \ref{corollary:MainCorollary}).
Finally, we show that if $\dpcore$ is a solution-preserving DP-core of table complexity $\alpha(k,n)$ solving a problem $\problem$, then given an $n$-vertex graph $G$  together with a tree decomposition $\treedec$ of width at most $\treewidthvalue$ and underlying tree structure $\treeT$, one can construct a $\treeT$-shaped tree automaton of width $\alpha(k,n)$ whose accepted terms encode solutions for $\problem$ in $\graphG$ (Theorem \ref{theorem:FromDPCoresToTreeAutomata}).
By combining Theorem \ref{theorem:FromDPCoresToTreeAutomata} with Corollary \ref{corollary:MainCorollary}, we infer that the extension complexity of the solution polytope of $\problem$ on $\graphG$ is upper bounded by $O(\alpha(k,n)\cdot n)$. Combining this result with the well known fact that linear programs are polynomial time solvable in the number of input inequalities,  we show that this conversion is optimal in Theorem \ref{OptimalConversion}.

\subsection{Generalizations}
Although for simplicity of exposition our main results will be stated in terms of vertex subset problems, it is worth noting that these results also generalize to edge subset problems, where the goal is to find a subset of edges satisfying a given condition (for instance, {\sc Cut$_{\ell}$}, {\sc HamiltonianCycle}), and to problems where feasible solutions are tuples of subsets of vertices, edges, or both. For instance for each fixed $d$, the {\sc $d$-coloring} problem is a problem where the goal is to partition the vertex set of the graph into $d$ subsets, each of which is an independent set.

\section{Preliminaries}
\subsection{Basic Definitions}
We denote by $\N$ the set of natural numbers and by $\Nplus$ the set of positive natural numbers. For each $n\in \N$, we let 
$[n] = \{1,\dots,n\}$. In particular, $[0] = \emptyset$. 
Given a set $\setS$, the set of subsets of $\setS$ is denoted by $\powerset{\setS}$, and the set of finite subsets of $\setS$ is denoted by $\finitesubsets{\setS}$. Given a function $f:S\rightarrow R$, and $S'\subseteq S$, we let $f(S') = \{f(s)\;|\; s\in S'\}$ be the image of $S'$ under $f$.

We define a \emph{graph} as a triple $\graphG = (\vertexset,\edgeset,\rho)$, where $\vertexset \subset \N$ is a finite set of \emph{vertices}, $\edgeset \subset \N$ is a finite set of \emph{edges} and $\rho \subset \edgeset \times \vertexset$ is an incidence relation. For an edge $\edgee \in \edgeset$, we let $\epoints{\edgee} = \{\vertexv \in \vertexset | (\edgee,\vertexv) \in \rho\}$ denote the set of vertices incident to $\edgee$. We may write $\vertexset_\graphG$, $\edgeset_\graphG$ and $\rho_\graphG$ to denote sets $\vertexset$, $\edgeset$ and $\rho$. We define the \emph{empty graph} as the graph $(\emptyset,\emptyset,\emptyset)$ with neither vertices, nor edges. We let \graphs denote the set of all graphs.

An \emph{isomorphism} from a graph $\graphG$ to a graph $\graphH$ is a pair of functions $\phi = (\phi_1, \phi_2)$ where $\phi_1: \vertexset_\graphG \rightarrow \vertexset_\graphH$ is a bijection from the vertex set of $\graphG$ to the vertex set of $\graphH$ and $\phi_2: \edgeset_\graphG \rightarrow \edgeset_\graphH$ is a bijection from the edge set of $\graphG$ to the edge set of $\graphH$ such that for each vertex $\vertexv \in \vertexset_\graphG$ and each edge $\edgee \in \edgeset_\graphG$, we have that 
$(\edgee,\vertexv) \in \increl_\graphG$ if and only if $(\phi_2(\edgee),\phi_1(\vertexv)) \in \increl_\graphH$. If there exists such a bijection, we say that graphs $\graphG$ and $\graphH$ are \emph{isomorphic} and denote this by $\graphG \sim \graphH$.

\subsection{Vertex Subset Problems}
\begin{definition}[Vertex Subset Problem]
    A \emph{vertex subset problem} is a subset $\problem \subset \graphs \times \finitesubsets{\N}$ where the following conditions are satisfied for each pair $(\graphG,\solsetX)\in \problem$: 
    \begin{enumerate}
        \item $X\subseteq \vertexset_{\graphG}$, and 
        \item for each graph $\graphH$ isomorphic to $\graphG$, and each isomorphism $\phi=(\phi_1,\phi_2)$ from $\graphG$ to $\graphH$, $(\graphH,\phi_1(\solsetX))\in \problem$. 
    \end{enumerate}
\end{definition}

Given a graph $\graphG$, we say that a subset $X\subseteq \vertexset_{\graphG}$ is a {\em solution for $\problem$} in $\graphG$ if $(\graphG,X)\in \problem$. 
The set of solutions for $\problem$ in $\graphG$ is denoted by 

\begin{equation}\label{equation:SolutionSet}
    \solutionset_{\problem}(\graphG) = \{\solsetX\;:\; (\graphG,\solsetX)\in \problem\}.
\end{equation}

For example, for each $\ell\in \Nplus$, {\sc IndependentSet}$_\ell$ denotes the vertex subset problem consisting of all pairs $(\graphG,\solsetX)$ where $\graphG$ is a graph and $\solsetX$ is an independent set of size at least $\ell$ in $\graphG$, that is, a subset of vertices where no pair of vertices is connected by an edge. Other prominent examples of vertex subset problems are {\sc VertexCover$_\ell$}, {\sc DominatingSet$_\ell$} and {\sc Clique$_\ell$}. 

\subsection{Solution Polytopes}
Let $\varset=\{\varx_1,\varx_2,\dots,\varx_n\}$ be a set of variables. A vector over $\varset$ is a function $\realvec:\varset\rightarrow \R$. 
We say that a vector $\realvec:\varset\rightarrow \R$ satisfies an inequality $\sum_i\alpha_ix_i \leq b$ if the inequality holds whenever each variable $x_i$ is replaced by the value $\realvec(x_i)$. That is to say, if $\sum_i\alpha_i\realvec(x_i)\leq b$. 
A vector $\realvec$ is a convex combination of a set of vectors 
$U = \{\realvec_1,\dots,\realvec_m\}$ if $\realvec = \alpha_1\realvec_1 + \alpha_2\realvec_2 + \dots + \alpha_m\realvec_m$ for some non-negative real numbers $\alpha_i$ such that $\sum_{i}\alpha_i= 1$. The {\em convex hull} of $U$ is the set $\convexhull(U)$ of all convex combinations of vectors in $U$. We note that  
$\convexhull(U)$ is a polytope, and as such can be defined as the set of solutions of a system of linear inequalities. 

Let $\graphG$ be an $n$-vertex graph, and let $X\subseteq \vertexset_{\graphG}$. We let $\hat{X}:\mathcal{X}\rightarrow \R$ denote the vector over variables $\mathcal{X} = \{x_v\;:\;v\in \vertexset_{\graphG}\}$ where for each $v\in \vertexset_{\graphG}$, 
$\hat{X}(x_v) = 1$ if and only if $v\in X$. Given a vertex subset problem $\problem$ and a graph $\graphG$, we let $\polytopeP_{\problem}(\graphG)$ be the polytope defined as the convex hull of vectors associated with solutions of $\problem$ in $\graphG$. More specifically, 
$\polytopeP_{\problem}(\graphG) = \convexhull(\{\hat{X}\;:\; X\in \solutionset_{\problem}(\graphG)\})$. 

\subsection{Addressed Trees}
An {\em addressed tree} is a four-tuple $\treeT = (\nodeset,\arcset,\rootr,\addressmap)$ 
denoting a rooted tree with nodes $\nodeset$, root $\rootr \in \nodeset$, arcs $\arcset\subseteq \nodeset\times\nodeset$ directed from the leaves towards the root, and arc-labeling function 
$\addressmap:\arcset\rightarrow \N$ that labels the arcs of $\treeT$ with numbers in such a way that for each $\nodeu\in \nodeset$, the arcs in the set
$\incomming(\nodeu) = \{(\nodev,\nodeu)\;|\; (\nodev,\nodeu)\in \arcset\}$ are injectively labeled with numbers from $\{1,\dots,|\incomming(\nodeu)|\}$. For each arc $(v,u)\in \incomming(\nodeu)$, we say that $v$ is the $i$-th child of $u$ if $\addressmap(v,u)=i$. We may write $\nodes{\treeT}$ to denote the set of nodes of $\treeT$, and $\rootfunction{\treeT}$ to denote the root of $\treeT$.

\section{Dynamic Programming Cores for Tree Decompositions}\label{section:DPCoresTreeDecompositions}
The formalism we use to express dynamic programming algorithms operates on edge-introducing tree decompositions. For a matter of consistency with the notation in the remainder of the paper, our definition of tree decompositions uses a slightly distinct notation than is typically used in the literature. 

We define an \emph{edge-introducing} tree decomposition of a graph $\graphG$
as a triple $\treedec = (\tree,\bagfunction,\edgefunction)$ where 
$\tree$ is an addressed tree, $\edgefunction:\edgeset(\graphG)\rightarrow \nodes{\tree}$ is an \emph{injective} function, and $\bagfunction:\nodes{\tree}\rightarrow 
\mathcal{P}(\vertexset_{\graphG})$ is a function satisfying the following conditions: 
\begin{enumerate}
    \item For each vertex $\vertexv\in \vertexset(\graphG)$, there is a node $\nodeu\in \nodes{\tree}$ such that $\vertexv\in \bagfunction(\nodeu)$; 
    \item for each edge $\edgee \in \edgeset(\graphG)$, 
    $\epoints{\edgee}\subseteq \bagfunction(\edgefunction(\edgee))$; 
    \item for each $\vertexv \in \vertexset(\graphG)$, the set $\{\nodeu \in \nodes{\tree}\; |\; \vertexv \in \bagfunction(\nodeu)\}$ induces a connected sub-tree of $\tree$. 
\end{enumerate}

The \emph{width} of $\treedec$ is defined as $\max_{\nodeu}|\bagfunction(\nodeu)|-1.$ The treewidth of a graph $\graphG$, denoted by $\treewidth{\graphG}$, is defined as the minimum width of an edge-introducing tree decomposition of $\graphG$.

An edge-introducing tree decomposition $\treedec$ is said to be {\em nice} if each node $\nodeu\in \nodes{\tree}$ has at most two children and the following conditions are satisfied: If $\nodeu$ has two children $\nodeu'$ and $\nodeu''$, then $\bagfunction(\nodeu) = \bagfunction(\nodeu') = \bagfunction(\nodeu'')$. In this case, $\nodeu$ is called a {\em join node}. If $\nodeu$ has a single child $\nodeu'$, then it either introduces a vertex $\vertexv$ (meaning that $\bagfunction(\nodeu)\backslash \bagfunction(\nodeu') = \{\vertexv\}$), or it forgets a vertex $\vertexv$ (meaning that $\bagfunction(\nodeu')\backslash \bagfunction(\nodeu) = \{\vertexv\}$), or it introduces an edge $\edgee$ (meaning that  $\bagfunction(\nodeu)= \bagfunction(\nodeu')$ and $\edgefunction(\edgee) = \nodeu$). 
If $\nodeu$ is a leaf node, or the root node, then $\bagfunction(\nodeu) = \emptyset$. In this work, we assume that edge-introducing tree decompositions are nice. This assumption is without loss of generality, since any tree decomposition of width $\treewidthvalue$  of a graph $\graphG$ can be transformed into a nice edge-introducing tree decomposition of $\graphG$ of width at most $\treewidthvalue$ in time $O(\treewidthvalue\cdot |\nodes{\tree}|)$ {Kloks1994Treewidth}. 

We formalize the notion of a dynamic programming algorithm operating on tree decompositions using the notion of a {\em dynamic programming core}. Our formalism is essentially equivalent in expressiveness as the notion of a DP-core introduced in {baste2022diversity} in the sense that dynamic programming algorithms developed in either model can be easily converted to each other with minor adaptations. 

\begin{definition}
A DP-core is a $6$-tuple $\dpcore$ whose components are specified as follows. 
\begin{enumerate}
    \item $\leafcore$ is a finite non-empty subset of $\{0,1\}^*$. 
    \item $\introvertexcore:\N\times \{0,1\}^* \rightarrow \finitesubsets{\{0,1\}^*}$
    \item $\introedgecore:\N\times \N\times \{0,1\}^*\rightarrow \finitesubsets{\{0,1\}^*}$.
    \item $\forgetvertexcore:\N\times \{0,1\}^*\rightarrow \finitesubsets{\{0,1\}^*}$. 
    \item $\joincore:\{0,1\}^*\times \{0,1\}^*\rightarrow \finitesubsets{\{0,1\}^*}$. 
    \item $\final: \{0,1\}^* \rightarrow \{0,1\}$. 
\end{enumerate}
\end{definition}

For each $S\subseteq \{0,1\}^*$, we define $\introvertexcore(\vertexv,S) = \bigcup_{\witness\in S} \introvertexcore(\vertexv,\witness)$. 
The extension of the remaining components of $\dpcore$ to subsets of strings is defined analogously. 

Intuitively, a DP-core $\dpcore$ is a specification of a dynamic programming algorithm that operates on edge-introducing tree decompositions of graphs. The algorithm processes such a tree decomposition $\treedec = (\tree,\bagfunction,\edgefunction)$ from the leaves towards the root, assigning to each node $\nodeu\in \nodes{\tree}$ a finite subset of strings $\dynamization(\nodeu)$ according to the following inductive process. 
\begin{enumerate}
    \item If $\nodeu$ is a leaf node of $\tree$, $\dynamization(\nodeu) = \leafcore$
    \item If $\nodeu$ is an internal node of $\tree$ with a unique child $\nodeu'$, then 
    \begin{enumerate}
        \item $\dynamization(\nodeu) = \introvertexcore(\vertexv,\dynamization(\nodeu'))$ if vertex $\vertexv$ is introduced at $\nodeu$. 
        \item $\dynamization(\nodeu) = \forgetvertexcore(\vertexv,\dynamization(\nodeu'))$ if vertex $\vertexv$ is forgotten at $\nodeu$.
        \item $\dynamization(\nodeu) = \introedgecore(\vertexv,\vertexv',\dynamization(\nodeu'))$ if an edge connecting $\vertexv$ and $\vertexv'$ is introduced
        at $\nodeu$. 
    \end{enumerate}
    \item If $\nodeu$ is a join node of $\tree$ with children $\nodeu'$ and $\nodeu''$ then $\dynamization(u) = \joincore(\dynamization(\nodeu'),\dynamization(\nodeu''))$. 
\end{enumerate}
We say that $\dpcore$ accepts the tree decomposition $\treedec$ if the set $\dynamization(\rootfunction{\tree})$ associated with the root node of $\tree$ has a final string. That is to say, $\witness\in \dynamization(\rootfunction{\tree})$ with $\final(\witness)=1$. 
\begin{definition}
    We say that $\dpcore$ solves a vertex subset problem $\problem$ if for each graph $\graphG$, and each tree decomposition $\treedec$ of $\graphG$, $\dpcore$ accepts $\treedec$ if and only if $\solutionset_{\problem}(\graphG)\neq \emptyset$. 
\end{definition}

We say that $\dpcore$ has {\em table complexity} $\localcomplexity(\treewidthvalue,n)$ if for each $n$-vertex graph $\graphG$, each tree decomposition $\treedec=(\tree,\bagfunction,\edgefunction)$ of $\graphG$ 
of width at most $\treewidthvalue$, and each node $\nodeu\in \nodes{\tree}$, the set $\dynamization(\nodeu)$ has at most $\localcomplexity(\treewidthvalue,n)$ elements. We remark here that we do not specify the particularities of the runtime complexity of obtaining sets $\dynamization(\nodeu)$ a DP-core associates to a node $\nodeu$ of a tree decomposition. Only the table complexity of a DP-core will be relevant in the results and theorems that follow. 
\\ 

\noindent{\bf A DP-core for Independent Set.}
As an illustration, we define below a DP-core $\dpcore$ for the problem {\sc IndependentSet$_{\ell}$}. The DP-core is modeled against the textbook dynamic programming algorithm solving {\sc IndependentSet$_{\ell}$} parameterized by treewidth {BlueBook}. 
It is enough to specify the operation of each component of the core on strings that encode pairs of the form $(S,c)$ where $S\subseteq \N$ and $c\in \N$. The particular encoding itself is not relevant, as long as it is fixed. We let $\dpcore$ be the DP-core whose components are defined as follows. 

\begin{itemize}
    \item $\leafcore = \{(\emptyset,0)\}$. 
    \item $\introvertexcore(v,(S,c)) = \{(S,c),(S\cup \{v\},c+1)\}$.
    
    \item $\introedgecore(v,v',(S,c)) =
    \left\{\begin{array}{l}
    \emptyset \mbox{ if $\{v,v'\}\subseteq S$} \\ 
    \{(S,c)\} \mbox{ otherwise.} 
    \end{array}\right.$ 

    \item $\forgetvertexcore(v,(S,c)) = \{(S\backslash \{v\},c)\}$. 
\end{itemize}

\begin{itemize}
    \item
    {\small 
    ${\joincore((S_1,c_1),(S_2,c_2))=}
    \left\{
    \begin{array}{l}
    \emptyset \mbox{ if $S_1\neq S_2$} \\ 
    \{(S_1,c_1+c_2 - |S_1|)\} \mbox{ otw.} 
    \end{array}\right.
    $
    }
    \item $\final((S,c)) = 1$ if and only if $c\geq \ell$. 
\end{itemize}

The following well known proposition implies that the
DP-core $\dpcore$
specified above solves {\sc IndependentSet$_{\ell}$}. Below, $\graphG_{\nodeu}$ means the subgraph of $\graphG$ induced by the set $\bigcup_{\nodet\in \nodes{\treeT_\nodeu}} \bagfunction(t)$, where $\treeT_\nodeu$ is the sub-tree of $\treeT$ rooted at $\nodeu$. 

\begin{proposition}[\cite{BlueBook} Ch. 7.3, \cite{KleinbergTardos2006} Ch. 10.4]\label{proposition:CorrectnessIndependentSetCore}
    Let $\graphG$ be a graph, $\treedec = (\treeT,\bagfunction,\edgefunction)$ be a tree decomposition of $\graphG$. Then, for each $\nodeu\in \nodes{T}$, and each pair $(S,c)\in \powerset{\N}\times \N$, $(S,c)$  belongs to $\dynamization(\nodeu)$ if and only if the graph $\graphG_{\nodeu}$ has an independent set $I$ of size at least $c$ such that
    $S = I\cap \bagfunction(\nodeu)$. 
\end{proposition} 

In particular, $\graphG$ has an independent set of size at least $\ell$ if and only if for some $\ell'\geq \ell$, the pair $(\emptyset,\ell')$ belongs to $\dynamization(\rootfunction{\treeT})$, and this happens if and only if $\treedec$ is accepted by $\dpcore$. We also note that by Proposition \ref{proposition:CorrectnessIndependentSetCore}, we have that if $\treedec$ has width $\treewidthvalue$ and $\graphG$ has $n$ vertices, then $|\dynamization(\nodeu)| \leq 2^{\treewidthvalue} \cdot |\vertexset_{\graphG}|$. Therefore, the table complexity of $\dpcore$ is upper bounded by 
$2^{\treewidthvalue}\cdot n$. Although very simple, the DP-core defined above is essentially optimal under SETH \citep{lokshtanov2018known}.

DP-cores for problems {\sc DominatingSet$_{\ell}$}, {\sc Cut$_{\ell}$}, {\sc HamiltonianCycle} and {\sc $d$-Coloring} are given in appendix \ref{DP-Cores}.

\section{The Polytope of a $\treeT$-Shaped Language}
Let $\Sigma$ be a set of symbols. A term over $\Sigma$ is a pair $\term = (\treeT,\lambda)$ where $\treeT$ is an addressed tree, and $\lambda:\nodes{\treeT}\rightarrow \Sigma$ is a function that labels each node of $\treeT$ with a symbol from $\Sigma$. We say that a term $\term$ is $\treeT$-shaped if $\term = (\treeT,\lambda)$ for some $\lambda$. Whenever clear from the context, we may write $\term(\nodeu)$ to denote the label $\lambda(\nodeu)$. We let $\Terms{\Sigma}$ denote the set of all terms over $\Sigma$. 

\subsection{$T$-Shaped Tree Automata}
Tree automata are a generalization of string automata that processes labeled trees instead of strings. In this work we will use tree automata to bridge dynamic programming algorithms and extended formulations. In particular, we will be concerned with a restricted version of tree automata where the states and transitions of the automaton itself can be organized in the shape of a tree. More specifically, the shape of the automaton will be specified by an addressed tree $\treeT$. For a complete exposition on tree automata we refer the reader to \citep{comon2007tata}.

Let $\treeT$ be an addressed tree. A $\treeT$-shaped tree automaton is a tuple $\automaton = (\Sigma,\stateset,\finalstateset,\Delta)$ 
where $\stateset = \bigcup_{\nodeu} \stateset_{\nodeu}$ is a set of states, partitioned into a collection of subsets 
$\{\stateset_{\nodeu}\}_{\nodeu\in \nodes{\treeT}}$, $\finalstateset\subseteq \stateset_{\rootfunction{\treeT}}$ is a set of final states, and
$\Delta = \bigcup_{\nodeu} \Delta_{\nodeu}$ is a set of transitions, partitioned into a collection of subsets $\{\Delta_{\nodeu}\}_{\nodeu\in \nodes{\treeT}}$ such that the following condition is satisfied: for each node $\nodeu\in \nodes{\treeT}$ with children $\nodeu_1,\dots,\nodeu_s$, $$\Delta_{\nodeu}\subseteq \stateset_{\nodeu_1}\times \dots \times \stateset_{\nodeu_s} \times \Sigma \times \stateset_{\nodeu}.$$ 

Let $\term = (\treeT,\lambda)$ be a term over $\Sigma$. A trace for $\term$ in $\automaton$ is a function $\trace:\nodes{\treeT}\rightarrow \stateset$ such that for each node $\nodeu\in \nodes{\treeT}$ with children $\nodeu_1,\dots,\nodeu_{s}$, the tuple 
$(\trace(\nodeu_1),\dots,\trace(\nodeu_s),\term(\nodeu),\trace(\nodeu))$ is a transition in $\Delta_{\nodeu}$. We say that $\trace$ is an {\em accepting trace} if $\trace(\rootfunction{\treeT})\in \finalstateset$. We say that $\term$ is accepted by $\automaton$ if there is an accepting trace $\trace$ for $\term$ in $\automaton$. The {\em language} of $\automaton$ is the set $\languagemap(\automaton)$ of all terms accepted by $\automaton$.

\subsection{The Polytope $\polytopeP(\automaton)$}
Let $\treeT$ be an addressed tree and $\Sigma$ be a set of symbols. We let 
$$\mainvarset(\treeT,\Sigma) = \{\mainvar_{\nodeu,a}\;|\; \nodeu\in \nodes{T}, a\in \Sigma\}$$ denote the set of variables associated with $\treeT$ and $\Sigma$. Given a $\treeT$-shaped term $\term\in \Terms{\Sigma}$, we let
$\htau:\mainvarset(\treeT,\Sigma)\rightarrow \{0,1\}$ be the 0/1 vector defined by setting, for each $\nodeu\in \nodes{\treeT}$ and each symbol $a\in \Sigma$, $\htau(x_{\nodeu,a})=1$ if $\term(\nodeu) = a$ and $\htau(x_{\nodeu,a})=0$ otherwise. Given a $\treeT$-shaped tree automaton $\automaton$, we define the polytope associated with $\automaton$, denoted by $\SlicePolytope{\automaton}$, as the convex hull of all vectors encoding terms accepted by $\automaton$: \\
$$\SlicePolytope{\automaton}=
\convexhull(\{\htau|\term \in \languagemap(\automaton)\}).$$

\subsection{An Extended Formulation for $\polytopeP(\automaton)$}\label{subsection:ExtendedFormulationPA}
A polytope $\polytopeP'$ is an extended formulation of a polytope $\polytopeP$ if $\polytopeP$ is a projection of $\polytopeP'$. 
 The next theorem, which is the main result of this section, states that polytopes associated with $\treeT$-shaped tree automata have extended formulations that can be defined by systems of linear inequalities whose size is linear on the number of states plus the number of transitions of the automata. 
 
\begin{theorem}\label{theorem:MainTheoremImplicit}
    Let $\automaton = (\Sigma,\stateset,\finalstateset,\Delta)$ be a $\treeT$-shaped tree automaton. The polytope $\SlicePolytope{\automaton}$ has an extended formulation with $\bigO{|\stateset|+|\Delta|}$ extension variables and 
    $\bigO{|\stateset| + |\treeT|\cdot|\Sigma|}$ 
    inequalities.
\end{theorem}

Theorem \ref{theorem:MainTheoremImplicit} above follows directly from Theorem \ref{theorem:MainTheorem} appearing later in this section.

We define the {\em width} of a $\treeT$-shaped tree automaton $\automaton$, denoted by $\width(\automaton)$, as the maximum size of a state cell: 
\[\width(\automaton) = \max_{\nodeu\in \nodes{\treeT}} |\stateset_{\nodeu}|.\] 
Since a $\treeT$-shaped tree automaton of width $w$ can have at most $w\cdot |\treeT|$ states, Theorem \ref{theorem:MainTheoremImplicit} immediately implies the following corollary: 

\begin{corollary}\label{corollary:MainCorollary}
    Let $\automaton = (\Sigma,\stateset,\finalstateset,\Delta)$ be a $\treeT$-shaped tree automaton of width $\width(\automaton)$. The polytope $\SlicePolytope{\automaton}$ has extension complexity $\bigO{|\treeT|\cdot(\width(\automaton) + |\Sigma|)}$.
\end{corollary}

In the remainder of this section we define an extended formulation for the polytope $\polytopeP(\automaton)$ satisfying the conditions of Theorem \ref{theorem:MainTheoremImplicit}. We start by specifying the extension variables, which we split into two sets. The first set, 
\[\statevarset(\automaton) = \{\statevar_{\nodeu,\stateq} | \nodeu \in \nodes{\treeT}, \stateq \in \stateset_\nodeu\},\]
has one variable $\statevar_{\nodeu,\stateq}$ for each node $\nodeu$ in $\nodes{\treeT}$ and each state $\stateq$ in $\stateset_\nodeu$. We call the elements of $\statevarset(\automaton)$ {\em state variables}. The second set, 
\[\transvarset(\automaton) = \{\transvar_{\nodeu,\delta} | \nodeu \in \nodes{\treeT}, \delta \in \Delta_\nodeu\},\]
has one variable 
$\transvar_{\nodeu,\delta}$ for each node $\nodeu$ in $\nodes{\treeT}$ and each transition $\delta$ in $\Delta_\nodeu$. We call the elements of $\transvarset(\automaton)$ {\em transition variables}. We let  $\variables(\automaton) = \mainvarset(\treeT,\Sigma) \cup \statevarset(\automaton) \cup \transvarset(\automaton)$ be the set of variables associated with $\automaton$. In what follows, we will refer to the sets $\mainvarset(\treeT,\Sigma)$, $\statevarset(\automaton)$ and $\transvarset(\automaton)$ simply as $\mainvarset$, $\statevarset$ and $\transvarset$ respectively. 

Given a transition $\delta = (\stateq_1,\dots,\stateq_{s},a,\stateq)$
in $\Delta$, we say that $\stateq$ is the {\em consequent state} of $\delta$, while for each $i\in [s]$, $\stateq_i$ is an {\em antecedent state} of $\delta$. We denote the consequent state
$\stateq$ by $\consequent{\delta}$ and the set of antecedent states $\{\stateq_1,\dots,\stateq_s\}$ by $\antecedent{\delta}$. We denote by $\symbolfunction{\delta}$ the symbol $\symbola$ of $\delta$.

Let $\transvar_{\nodeu,\delta}$ be a transition variable in
$\transvarset$ for some $\nodeu\in \nodes{\treeT}$ and transition $\delta \in \transitionset_{\nodeu}$. If $\stateq = \consequent{\delta}$, we say that $\transvar_{\nodeu,\delta}$ is an \emph{incoming} transition variable of $\statevar_{\nodeu,\stateq}$. If $\nodeu'$ is a child of $\nodeu$ and $\stateq'\in \antecedent{\delta}$, we say that $\transvar_{\nodeu,\delta}$ is an \emph{outgoing} transition variable of $\statevar_{\nodeu',\stateq'}$. 

\begin{definition}\label{definition:VectorOfATrace}
Let $\automaton = (\Sigma,\stateset,\finalstateset,\Delta)$ be a $\treeT$-shaped tree automaton, $\term = (\treeT,\lambda)$ be a term in $\languagemap(\automaton)$, and $\trace:\nodes{\treeT}\rightarrow \stateset$
be a trace for $\term$ in $\automaton$. We let $\tracehat$ be the 0/1 vector over $\variables(\automaton)$ satisfying the following conditions: 

\begin{enumerate}
    \item\label{traceOne} For each $\mainvar_{\nodeu,\symbola} \in \mainvarset$, $\tracehat(\mainvar_{\nodeu,\symbola}) = 1$ iff $\lambda(\nodeu)=\symbola$. 
    \item\label{traceTwo} For each $\statevar_{\nodeu,\stateq}\in \statevarset$, $\tracehat(\statevar_{\nodeu,\stateq})=1$ iff $\trace(\nodeu) = \stateq$. 
    \item\label{traceThree} For each $\transvar_{\nodeu,\delta}$ in $\transvarset$, $\tracehat(\transvar_{\nodeu,\delta}) = 1$ iff 
    $\delta =(\trace(\nodeu_{1}),\dots,\trace(\nodeu_{s}),\lambda(\nodeu), \trace(\nodeu))$, where $\nodeu_1,\dots,\nodeu_s$ are the children of $\nodeu$. 
\end{enumerate}
\end{definition}

Let $\term$ be a term in $\languagemap(\automaton)$, and $\trace$ be a trace for $\term$ in $\automaton$. Then by restricting the vector $\tracehat$ to the variables in $\mainvarset$, we obtain the vector $\htau$ associated with $\term$. This implies that the polytope 
\begin{equation}\label{equation:Pex}
    \polytopex{\automaton} = \convexhull(\{\tracehat|\trace \mbox{ is an accepting trace in $\automaton$}\})
\end{equation}
defined as the convex-hull of vectors corresponding to {\em accepting} traces in $\automaton$ is an extended formulation of the polytope $\polytopeP(\automaton)$. Note that $\statevarset$ has $|\stateset|$ variables, while the set $\transvarset$ has $|\Delta|$ variables. Therefore, besides the main variables in $\mainvarset$, we are using $|\stateset|+|\Delta|$ extension variables to define $\polytopex{\automaton}$.

Next, we will show that $\polytopex{\automaton}$ can be defined as the set of solutions of a system of linear 
inequalities
$\mip_\automaton$
containing $O(|\stateset| + |\treeT|\cdot|\Sigma|)$ linear constraints. 
The first set of constraints ensure that all variables in a solution should assume values between $0$ and $1$. More specifically, for each main variable $\mainvar_{\nodeu,\symbola}\in \mainvarset$, each state variable $\statevar_{\nodeu,\stateq}\in \statevarset$, and each transition variable $\transvar_{\nodeu,\delta}\in \transvarset$, 

\begin{equation}\label{eq:2}
    0\leq \mainvar_{\nodeu,\symbola} \leq 1 \hspace{0.5cm}
    0\leq \statevar_{\nodeu,\stateq} \leq 1 \hspace{0.5cm}
    0\leq \transvar_{\nodeu,\delta} \leq 1
\end{equation}

Let $\rootr = \rootfunction{\treeT}$ be the root of $\treeT$. The following constraints ensure that in an integral solution, there is exactly one final state $\stateq\in \finalstateset$ for which variable $\statevar_{\rootr,\stateq}$ is set to $1$. Additionally, in such a solution, for each state $\stateq \in \stateset_{\rootr}\backslash \finalstateset$ the variable $\statevar_{\rootr,\stateq}$ is set to $0$.

\begin{equation}\label{finalState}
    \sum_{\stateq \in \finalstateset} \statevar_{\rootr,\stateq}=1
\end{equation}
\begin{equation}\label{nonFinalState}
    \forall \stateq\in \stateset_{r}\backslash \finalstateset:\ \  \statevar_{r,q}=0 
\end{equation}

The next set of constraints ensure that for 
each node $\nodeu\in \nodes{\treeT}$, and each state $\stateq\in \stateset_\nodeu$, for any integral solution that sets variable $\statevar_{\nodeu,\stateq}$ to $1$ there is some transition $\delta\in \transitionset_{\nodeu}$ with consequent $\stateq$ such that $\transvar_{\nodeu,\delta}$ is set to $1$. 

\begin{equation}\label{transvarConsequent}
    \forall \nodeu\in \nodes{\treeT}, \forall \stateq\in \stateset_\nodeu\!: \ \ \sum_{\delta, \consequent{\delta}=\stateq}\transvar_{\nodeu,\delta} = \statevar_{\nodeu,\stateq}
\end{equation}

The next set of constraints ensure that for 
each node $\nodeu\in \nodes{\treeT}$ distinct from the root $\rootr$, and each state $\stateq\in \stateset$, for any integral solution that sets variable $\statevar_{\nodeu,\stateq}$ to $1$ there is some transition $\transvar_{\parent{\nodeu},\delta}\in \transitionset_{\parent{\nodeu}}$ having $\stateq$ as an antecedent such that $\transvar_{\parent{\nodeu},\delta}$ is set to $1$. Here, $\parent{\nodeu}$ denotes the parent of $\nodeu$. 

\begin{equation}\label{transvarAntecedent}
    \forall \nodeu\in \nodes{\treeT}\backslash \{r\}, \forall \stateq\in \stateset_\nodeu\!: \ \ \sum_{\delta, \stateq \in \antecedent{\delta}}\transvar_{\parent{\nodeu},\delta} = \statevar_{\nodeu,\stateq}
\end{equation}

Finally, the next set of equations ensure that
if an integral solution sets a main variable $\mainvar_{\nodeu,\symbola}\in \mainvarset$ to $1$, then there is some transition 
$\delta\in \transitionset_{\nodeu}$ with 
$\symbolfunction{\delta}=\symbola$ such that
the transition variable $\transvar_{\nodeu,\delta}$ is also 
set to $1$. 

\begin{equation}\label{transvarSymbol}
    \forall \nodeu\in \nodes{\treeT}, \forall \symbola\in \Sigma\!: \sum_{\delta,\symbolfunction{\delta}=\symbola}\transvar_{\nodeu,\delta} = \mainvar_{\nodeu,\symbola}
\end{equation}

Theorem \ref{theorem:MainTheoremImplicit} is a direct consequence of the following theorem, stating that the set of inequalities $\mip_{\automaton}$ defines the polytope $\polytopex{\automaton}$.  

\begin{theorem}
\label{theorem:MainTheorem}
Let $\automaton = (\Sigma,\stateset,\finalstateset,\Delta)$ be a $\treeT$-shaped tree automaton. A vector $\mu$ belongs to $\polytopex{\automaton}$ if and only if $\mu$
satisfy all inequalities in $\mip_\automaton$. 
\end{theorem}

The proof of Theorem \ref{theorem:MainTheorem} is carried out in detail in the next subsection. The proof is split into three parts. In the first part, we show that for each trace $\trace$ in $\automaton$, the 0/1- vector $\tracehat$ satisfies the system of linear inequalities $\mip_{\automaton}$. This implies that the polytope $\polytopex{\automaton}$ is contained in the polytope $\polytopeP(\mip_{\automaton})$ defined by $\mip_{\automaton}$. In the second part, we show that if $\mu$ is an integral 
vector satisfying all inequalities of $\mip_{\automaton}$, then there is some trace 
$\trace$ of $\automaton$ such that $\tracehat = \mu$. This shows that at least the integral solutions of $\mip_{\automaton}$ are contained in $\polytopex{\automaton}$. Finally, in the third part, we show that any vertex of the polytope defined by $\mip_{\automaton}$ is a $0/1$-vector. Therefore, since $0/1$-vectors cannot be convex combinations of other $0/1$ vectors, we have that the vertices of $\polytopeP(\mip_{\automaton})$ are precisely the $0/1$-vectors satisfying all constraints in $\mip_{\automaton}$. This shows that $\polytopeP(\mip_{\automaton}) = \polytopex{\automaton}$, and concludes the proof of Theorem \ref{theorem:MainTheorem}.  

\subsection{Proof of Theorem \ref{theorem:MainTheorem}}
In this section, we prove Theorem \ref{theorem:MainTheorem}, which states that the set of linear inequalities defined in Subsection \ref{subsection:ExtendedFormulationPA} specifies the polytope $\polytopex{\automaton}$. The proof is split into three main parts, each of which will be covered separately. 

\subsubsection{Traces correspond to integral solutions.}
In this subsection, we show that for each trace $\trace$ in $\automaton$, the 0/1- vector $\tracehat$ satisfies the system of linear inequalities $\mip_{\automaton}$ (Lemma \ref{lemma:partI}). This implies that the polytope $\polytopex{\automaton}$ is contained in the polytope $\polytopeP(\mip_{\automaton})$ defined by $\mip_{\automaton}$.

\begin{lemma}
\label{lemma:partI}
 Let $\automaton$ be $\treeT$-shaped automaton, $\term$ be a term in $\languagemap(\automaton)$, and $\trace$ be a trace for $\term$ in $\automaton$. Then $\tracehat$ satisfies $\mip_{\automaton}$. 
 \end{lemma}
 \begin{proof}
 We will show that the vector $\tracehat$ satisfies inequalities (\ref{finalState})-(\ref{transvarSymbol}) defined in Subsection \ref{subsection:ExtendedFormulationPA}. Inequalities (\ref{eq:2}) are trivially satisfied by the definition of $\tracehat$.
 \begin{enumerate}
     \item Let $\rootr = \rootfunction{\treeT}$. Since $\trace$ is an accepting trace, there is some final state $\stateq\in \finalstateset$ such that $\trace(\rootr) = \stateq$. Therefore, by Condition \ref{traceTwo} of Definition \ref{definition:VectorOfATrace},  $\tracehat(\statevar_{\rootr,\stateq}) = 1$, while for every $\stateq'\in \stateset_{\rootr}\backslash \{\stateq\}$, we have that $\tracehat(\statevar_{\rootr,\stateq'})=0$. This implies that inequalities (\ref{finalState}) and (\ref{nonFinalState}) are satisfied. 
     \item Now, from Condition \ref{traceOne} of Definition \ref{definition:VectorOfATrace}, for each $\nodeu\in \nodes{\treeT}$, we have that $\tracehat(\mainvar_{\nodeu,\term(\nodeu)}) = 1$ while for every symbol $\symbola\in \Sigma\backslash \{\term(\nodeu)\}$, $\tracehat(\mainvar_{\nodeu,\symbola})=0$. 
     Similarly from Condition \ref{traceThree}, we have that for $\delta = (\trace(\nodeu_1),\dots,\trace(\nodeu_s),\term(\nodeu),\trace(\nodeu))$, $\tracehat(\transvar_{\nodeu,\delta})=1$, while for every other transition $\delta' \in \Delta_u \backslash \{\delta\}$, $\tracehat(\transvar_{\nodeu,\delta'}) = 0$. This implies that inequalities (\ref{transvarSymbol}) are satisfied.
     \item By Condition \ref{traceTwo} of Definition \ref{definition:VectorOfATrace}, we have that for each $\nodeu\in \nodes{\treeT}$, $\statevar_{\nodeu,\trace(\nodeu)}$ is the only state variable corresponding to node $\nodeu$ such that $\tracehat(\statevar_{\nodeu,\trace(\nodeu)})=1$. 
     Additionally, let $\delta=(\trace(\nodeu_1),\dots,\trace(\nodeu_s),\term(\nodeu), \trace(\nodeu))$. Then, by Condition \ref{traceThree} of Definition \ref{definition:VectorOfATrace}, $\transvar_{\nodeu,\delta}$ is the only variable corresponding to node $\nodeu$ such that $\tracehat(\transvar_{\nodeu,\delta})=1$. Since $\transvar_{\nodeu,\delta}$ is the incoming transition variable of $\statevar_{\nodeu,\trace(\nodeu)}$, inequalities (\ref{transvarConsequent}) are satisfied. 
    \item Finally, by Condition \ref{traceThree} of Definition \ref{definition:VectorOfATrace}, for each $\nodeu\in \nodes{\treeT}\backslash \{r\}$, there is a unique outgoing transition variable $\transvar_{\parent{\nodeu},\delta}$ of state variable $\statevar_{\nodeu,\trace(\nodeu)}$ that is set to $1$ by $\tracehat$ whenever state variable  $\statevar_{\nodeu,\trace(\nodeu)}$ is set to $1$. Therefore, 
    inequalities (\ref{transvarAntecedent}) are also satisfied.
 \end{enumerate}
 \end{proof}

\subsubsection{Integral Solutions Correspond to Traces}
Next, we show that if $\mu$ is an integral vector satisfying all inequalities of $\mip_{\automaton}$, then there is some trace 
$\trace$ of $\automaton$ such that $\tracehat = \mu$ (Lemma \ref{vectortoterm}). This shows that at least the integral solutions of $\mip_{\automaton}$ are contained in $\polytopex{\automaton}$. Before stating Lemma \ref{vectortoterm}, we prove the following useful lemmas and corollaries. 

In lemma \ref{EntriesLessThan1} we perform induction on the depth of vectors representing terms of a $\treeT$-shaped tree automaton. 
Let $\automaton = \{\Sigma,\stateset,\finalstateset,\Delta\}$ be a $\treeT$ shaped tree automaton. For each node $\nodeu \in \nodes{\treeT}$ we define the following subsets of $\statevarset$, $\transvarset$ and $\mainvarset$:
$$ \statevarset_\nodeu = \{\statevar_{\nodeu,\stateq} : \stateq \in \stateset_\nodeu\}$$
$$ \transvarset_\nodeu = \{\transvar_{\nodeu,\delta} : \delta \in \Delta_\nodeu\}$$
$$ \mainvarset_\nodeu = \{\mainvar_{\nodeu,\symbola} : \symbola \in \Sigma\}$$
We denote the union of theses sets for some node $\nodeu$ by $\ancvars_\nodeu = \statevarset_\nodeu \cup \transvarset_\nodeu \cup \mainvarset_\nodeu$ and we note that $\bigcup_\nodeu \statevarset_\nodeu = \statevarset$, $\bigcup_\nodeu \transvarset_\nodeu = \transvarset$ and $\bigcup_\nodeu\mainvarset_\nodeu = \mainvarset$. Moreover $\bigcup_\nodeu \ancvars_\nodeu = \variables(\automaton)$. 
For any vector $\eta$ over $\variables(\automaton)$, we write $\eta_{\ancvars_\nodeu}$ to denote the assignment of values to variables in $\ancvars_\nodeu$ according to the entries of $\eta$.

 \begin{lemma} \label{SumEntriesLessThan1}
    Let $\automaton$ be a $\treeT$-shaped tree automaton and let $\vectormu$ be a solution to $\mip_\automaton$. For each $\nodeu \in \nodes{\treeT}$ the sum of all variables in $\statevarset_\nodeu$ is equal to $1$, the sum of all variables in $\transvarset_\nodeu$ is equal to $1$ and the the sum of all variables in $\mainvarset_\nodeu$ is equal to $1$ under assignment $\vectormu_{\ancvars_\nodeu}$. 
\end{lemma}
\begin{proof}
    The proof is by induction on the depth a node of $\treeT$. 

    \noindent \textbf{Basis step:} We show first that the claim holds for sets $\statevarset_\rootr$, $\transvarset_\rootr$ and $\mainvarset_\rootr$ where $\rootr$ is the root of $\treeT$. Since $\vectormu$ is a solution, equations (\ref{finalState}) and (\ref{nonFinalState}) are satisfied and so the claim holds for $\statevarset_\rootr$, that is to say $$\sum_{\stateq \in \stateset_\rootr}\statevar_{\nodeu,\stateq} = 1.$$
    
    By definition, the transitions in $\Delta_\rootr$ are exactly those transitions with consequent in $\stateset_\rootr$. Additionally each transition has a unique consequent. Therefore, by equation (\ref{transvarConsequent}), the previous sum of state variables can be rewritten as follows
    
    \begin{equation}\label{sum:SumTransvarsByState}
        \sum_{\delta, \consequent{\delta}=\stateq_1}\transvar_{\rootr,\delta} + \sum_{\delta, \consequent{\delta}=\stateq_2}\transvar_{\rootr,\delta} + \dots + \sum_{\delta, \consequent{\delta}=\stateq_{|\stateset_\rootr|}}\transvar_{\rootr,\delta} = 1
    \end{equation}
    
    where $\stateq_i \in \stateset_\rootr$. This shows that the claim holds for $\transvarset_\rootr$.

    Further, since each transition has a unique symbol, the sum of variables in $\transvarset_\rootr$ can be rearranged as,

    \begin{equation}\label{sum:SumTransvarsBySymb}
        \sum_{\delta, \symbolfunction{\delta}=\symbola_1}\transvar_{\rootr,\delta} + \sum_{\delta, \symbolfunction{\delta}=\symbola_2}\transvar_{\rootr,\delta} + \dots + \sum_{\delta, \symbolfunction{\delta}=\symbola_{|\Sigma|}}\transvar_{\rootr,\delta}
    \end{equation}

    for $\symbola_i \in \Sigma$. Since $\vectormu$ is a solution it satisfies equations (\ref{transvarSymbol}) and therefore sum ($\ref{sum:SumTransvarsBySymb}$) can be rewritten as \begin{equation}\label{sum:SumMainVars}
    \sum_{\symbola \in \Sigma}\mainvar_{\nodeu,\symbola} = 1.
    \end{equation}  
    This shows that the claim holds for $\mainvarset_\rootr$.

    \noindent \textbf{Inductive step:} Let $\nodeu$ be an arbitrary node of $\treeT$ and assume that the claim  holds for parent $\parent{u}$ of $\nodeu$. We show first that the sum of all variables in $\statevarset_\nodeu$ is equal to 1. Since $\vectormu$ is a solution, by equation (\ref{transvarAntecedent}) state variable $\statevar_{\nodeu,\stateq}$ can be expressed as the sum of all transition variables of $\transvarset_{\parent{u}}$ for which $\statevar_{\nodeu,\stateq}$ is outgoing. Since by definition each transition $\delta$ takes a single antecedent $\stateq$ from $\stateset_\nodeu$, it follows that any two distinct state variables in $\statevarset_\nodeu$ cannot be outgoing for the same transition variable in $\transvarset_{\parent{u}}$. In other words, for any given $\transvar_{\parent{u},\delta}$,  it cannot be the case that both $\stateq \in \antecedent{\delta}$ and $\stateq' \in \antecedent{\delta}$ for $\stateq, \stateq' \in \stateset_\nodeu$. Therefore, the sum of transition variables in $\transvarset_{\parent{\nodeu}}$ can be written as the following sum of sums,

    \begin{equation}\label{sum:SumTransvarsParents}
        \sum_{\delta, \stateq_1 \in \antecedent{\delta}}\transvar_{\parent{\nodeu},\delta} + \sum_{\delta, \stateq_2 \in \antecedent{\delta}}\transvar_{\parent{\nodeu},\delta} + \dots + \sum_{\delta, \stateq_{|\stateset_\nodeu|} \in \antecedent{\delta}}\transvar_{\parent{\nodeu},\delta}
    \end{equation}

    where $\stateq_i \in \stateset_\nodeu$. By the equation (\ref{transvarAntecedent}) sum (\ref{sum:SumTransvarsParents}) is equal to sum of all state variables in $\statevarset_\nodeu$ and by the induction hypothesis sum (\ref{sum:SumTransvarsParents}) equals 1.

    The fact that the claim holds for sets $\transvarset_\nodeu$ and $\mainvarset_\nodeu$ can be shown by the same argument used for the base case by replacing $\rootr$ with $\nodeu$ in sums (\ref{sum:SumTransvarsByState}) and (\ref{sum:SumTransvarsBySymb}) respectively.
\end{proof}

The following corollary follows immediately form Lemma \ref{SumEntriesLessThan1}.

\begin{corollary}\label{EntriesLessThan1}
    Let $\automaton$ be a $\treeT$-shaped tree automaton and let $\vectormu$ be a solution to $\mip_\automaton$. Then all entries of $\vectormu$ are less than or equal to 1.
\end{corollary} 

If we impose integrality on $\vectormu$ we can derive the following corollary of Lemma \ref{SumEntriesLessThan1}.

\begin{corollary}\label{uniquevarvalued1}
    Let $\automaton$ be a $\treeT$-shaped tree automaton and let $\vectormu$ be a 0/1-vector
    over $\variables(\automaton)$ satisfying the system of inequalities $\mip_\automaton$. For each $\nodeu \in \nodes{\treeT}$, there is exactly one symbol $\symbola_{\nodeu}\in \Sigma$ such that $\vectormu(\mainvar_{\nodeu,\symbola_{\nodeu}})=1$, 
    exactly one state $\stateq_{\nodeu}\in \stateset_{\nodeu}$  such that $\vectormu(\statevar_{\nodeu,\stateq_{\nodeu}})=1$, and exactly one transition $\delta_{\nodeu}\in \Delta_{\nodeu}$ such that $\vectormu(\transvar_{\nodeu,\delta_{\nodeu}})=1$. 
\end{corollary}
\begin{proof}
    For each $\nodeu$ the sum over entries in each set $\statevarset_\nodeu$, $\transvarset_\nodeu$ and $\mainvarset_\nodeu$ must equal to $1$ by Lemma \ref{SumEntriesLessThan1}. Therfore, exactly one variable in each of sets $\statevarset_\nodeu$, $\transvarset_\nodeu$ and $\mainvarset_\nodeu$ can be non-zero (i.e. equalt to $1$) under $\vectormu$. 
\end{proof}

Using Corollary \ref{uniquevarvalued1}, we will show that from each integral solution to $\mip_{\automaton}$ one can extract a term $\term\in \languagemap(\automaton)$ together with an accepting trace $\trace$ for $\term$ in $\automaton$. 

\begin{lemma}\label{vectortoterm} 
    Let $\automaton$ be a $\treeT$-shaped tree automaton and let vector $\vectormu$ be a 0/1-solution to $\mip_{\automaton}$. Then there is some term $\term\in \languagemap(\automaton)$, and some accepting trace $\rho$ for $\term$ in $\automaton$ such that $\vectormu = \tracehat$. 
\end{lemma}
\begin{proof}
    Let $\vectormu$ be a 0/1-vector
    over $\variables(\automaton)$ satisfying the system of inequalities $\mip_\automaton$.
    By Corollary \ref{uniquevarvalued1}, for each $\nodeu \in \nodes{\treeT}$, there is
    \begin{itemize}
        \item exactly one symbol $\symbola_{\nodeu}\in \Sigma$ such that $\vectormu(\mainvar_{\nodeu,\symbola_{\nodeu}})=1$, 
        \item exactly one state $\stateq_{\nodeu}\in \stateset_{\nodeu}$  such that $\vectormu(\statevar_{\nodeu,\stateq_{\nodeu}})=1$, and
        \item exactly one transition $\delta_{\nodeu}\in \Delta_{\nodeu}$ such that $\vectormu(\transvar_{\nodeu,\delta_{\nodeu}})=1$.
    \end{itemize}
    Let $\term$ be the term defined by setting $\term(\nodeu) = \symbola_{\nodeu}$ for each node $\nodeu\in \nodes{\treeT}$. Additionally, let $\trace:\nodes{\treeT}\rightarrow \stateset$ be the function that sets 
    $\trace(\nodeu) = \stateq_{\nodeu}$ for each $\nodeu\in \nodes{\treeT}$. Now let $\nodeu\in \nodes{\treeT}$, be a node with children $\nodeu_1,\dots,\nodeu_s$. 
    Then, inequalities (\ref{transvarConsequent}) guarantee that $\stateq_\nodeu$ is the consequent state of $\delta_{\nodeu}$. On the other hand, inequalities (\ref{transvarAntecedent}) guarantee that for each $i\in \{1,\dots,s\}$, $\stateq_{\nodeu_i}$ is an antecedent of $\delta_{\nodeu}$. This implies that 
    $\delta_{\nodeu}$ is equal to $(\stateq_{\nodeu_1},\dots,\stateq_{\nodeu_s},\symbola_{\nodeu},\stateq_{\nodeu})$. This shows that $\trace$ is a trace for $\term$ in $\automaton$, and that $\tracehat = \vectormu$. Additionally, let $\rootr$ be the root of $\term$. Then by inequalities (\ref{finalState}), $\stateq_\rootr \in \finalstateset$ and $\vectormu(\statevar_{\rootr,\stateq_\rootr}) = 1$. Therefore $\trace$ is an accepting trace for $\term$ in $\automaton$.
\end{proof}

\subsubsection{Vertices of $\polytopex{\automaton}$ are integral}
Finally, in this subsection we show that any vertex of the polytope defined by $\mip_{\automaton}$ is a $0/1$-vector (Lemma \ref{lemma:NoNonintegralSolution}). Therefore, since $0/1$-vectors cannot be convex combinations of other $0/1$ vectors (Observation \ref{IntSolsAreVerts}), we have that the vertices of $\polytopeP(\mip_{\automaton})$ are precisely the $0/1$-vectors satisfying all constraints in $\mip_{\automaton}$. In other words, 
$\polytopeP(\mip_{\automaton})$ is the convex-hull of integral points satisfying $\mip_{\automaton}$. 
Since by Lemma \ref{vectortoterm}, for each integral solution $\vectormu$ to $\mip_{\automaton}$ there is an accepting trace $\trace$ such that $\tracehat = \vectormu$, we have that $\polytopeP(\mip_{\automaton}) = \polytopex{\automaton}$. This will conclude the proof of Theorem \ref{theorem:MainTheorem}.  

The following is a standard definition of extreme points (vertices) of a convex set, rewritten in the context of convex polytopes \citep{TroppConvexNotes, CornellExtremePointNotes, MuellerHalleConvex}. 

\begin{definition} \label{VerticesCannotBeDecomoposed}
    Let $P$ be a convex polytope. A point $\vectormu \in P$ is a vertex of $P$ if and only if there are no two distinct points $x,y \in P$ such that $\vectormu$ is a convex combination of $x$ and $y$.
\end{definition} 

From Corrollary~\ref{EntriesLessThan1} an the definition above it follows that integral solutions of $\mip_{\mathcal A}$ correspond to vertices of polytope $\polytopeP(\mip_\automaton)$.  The next observation Formalizes this fact. 

\begin{observation}\label{IntSolsAreVerts}
    Let $\mu$ be an integral solution to $\mip_{\mathcal A}$. 
    Then $\mu$ is a vertex of the polytope $\polytopeP(\mip_\automaton)$.
\end{observation}
\begin{proof}
    Suppose, for a contradiction, that $\mu$ is not a vertex of $\polytopeP(\mip_\automaton)$. 
    Then there exist distinct vertices $\mu_1,\dots,\mu_r$ of $\polytopeP(\mip_\automaton)$ and coefficients 
    $\lambda_1,\dots,\lambda_r\ge 0$ with $\sum_{i=1}^r \lambda_i=1$ and at least two $\lambda_i>0$ such that
    \[
    \mu \;=\; \sum_{i=1}^r \lambda_i \mu_i.
    \]
    By Corollary~\ref{EntriesLessThan1}, all entries of $\mu$ and of each $\mu_i$ lie in $[0,1]$. 
    
    We now use the following elementary fact from polyhedral and linear programming theory essentially stating that the 0/1 vectors of the $d$ dimensional hypercube are vertices (see example 0.4 of \cite{Ziegler1995}).
    
    \emph{Fact.}  If $x\in\{0,1\}^d$ is a convex combination of vectors $y_1,\dots,y_r$, and each $y_i\in[0,1]^d$, then for every $i$ with $\lambda_i>0$ we have $y_i=x$.
    
    Because at least two coefficients $\lambda_i$ are positive, the above Fact implies $\mu_i=\mu$ for all $i$ with $\lambda_i>0$, which contradicts the assumption that the $\mu_i$ are distinct. 
    Hence $\mu$ must be a vertex of $\polytopeP(\mip_\automaton)$.
    \end{proof}
    
    It now remains only to show that any non-integral solution of $\polytopeP(\mip_\automaton)$ is not a vertex.
    
    Given a vector $y\in \R^n$, we let the support of $y$ be the set of all coordinates where $y$ is non-zero. The size of the support of $y$ is the number of coordinates in it. 
    
\begin{observation}\label{SupportAndAlpha}
    Let $\automaton$ be a  $\treeT$-shaped tree automaton, let $\vectormu$ be a non integral solution to $\mip_\automaton$ with support set $S$ and let $\vectormu^*$ be an integral solution with support set $S^*$. If $S^*$ is contained in $S$ then,
        
    \begin{enumerate}
        \item\label{alphalessthan1} $0< \alpha< 1$ for $\alpha = \min\{\vectormu(i)\;:\; i\in S^*\}$
        \item\label{SprimeinS} $S^*$ is strictly contained in $S$
    \end{enumerate} 
\end{observation}
\begin{proof}
    Since $\vectormu$ and $\vectormu^*$ are solutions, we have that for each of these vectors, and each node $\nodeu$, the sum of coordinates corresponding to each of the sets $\statevarset_{\nodeu}$, $\transvarset_{\nodeu}$ and $\mainvarset_{\nodeu}$ is equal to $1$ (Lemma \ref{SumEntriesLessThan1}). For integral solution $\vectormu^*$ we now further that exactly one variable is sets, $\statevarset_{\nodeu}$, $\transvarset_{\nodeu}$ and $\mainvarset_{\nodeu}$ is equal to $1$ by Corollary \ref{uniquevarvalued1}. However since the sum over the coordinates over the selected set is already equal to $1$ under $\vectormu^*$, we have that for at least one such set, say $\statevarset_{\nodeu}$ without loss of generality, there must exist a coordinate in $\statevarset_{\nodeu}$ taking value less than $1$ with respect to $\vectormu$ (otherwise $\vectormu$ and $\vectormu^*$ would be equal). This shows that the smallest non-zero entry $\alpha$ in $\vectormu$ chosen among the support set of $\vectormu^*$,  $\alpha = \min\{\vectormu(i)\;:\; i\in S^*\}$ is less than $1$. Further, $\alpha>0$ because it is chosen from the support set of $\vectormu$, which only contains positive coordinates. This completes the proof for point \ref{alphalessthan1} of the Observation.
    
    To see that point \ref{SprimeinS} holds, assume without loss of generality that the coordinate, $\arg\min\{\vectormu(i)\;:\; i\in S^*\}$, taking value $\alpha$ with respect to $\vectormu$ occurs in set $\statevarset_{\nodeu}$. In order for the sum over entries in $\statevarset_{\nodeu}$ to equal to $1$ (as per Lemma \ref{SumEntriesLessThan1}) there must exist at least one other non-zero coordinate in $\statevarset_{\nodeu}$. This shows that $S^*$ is strictly contained in $S$. 
\end{proof}

\begin{lemma}\label{lemma:SubsolutionIntegral}
    Let $\automaton = \{\Sigma,\stateset,\finalstateset,\Delta\}$ be a $\treeT$-shaped tree automaton and let $\vectormu$ be a non-integral solution to  $\mip_\automaton$. Then, there exists a 0/1-solution $\vectormu^*$ whose support is strictly contained in the support of $\vectormu$. 
\end{lemma}
\begin{proof}
    We construct vector $\vectormu^*$ inductively from the entries of $\vectormu$ and observe that for each $\nodeu \in \treeT$ vector $\vectormu^*$ satisfies inequalities (\ref{eq:2}) through (\ref{transvarSymbol}). We start by instantiating variables in $\ancvars_\rootr$ where $\rootr = \rootfunction{\treeT}$. 
    
    \textbf{Basis step: }
    
    We choose arbitrarily a state variable $\statevar_{\rootr,\stateq}$ in $\statevarset_\rootr$ for which entry $\vectormu(\statevar_{\rootr,\stateq})$ is non-zero (at least one such variable must exist otherwise vector $\vectormu$ would not satisfy equation \ref{finalState}) and set the corresponding entry, $\vectormu^*(\statevar_{\rootr,\stateq})$, in vector $\vectormu^*$ to 1. Further, we select arbitrarily an incoming transition variable $\transvar_{\rootr,\delta} \in \transvarset_\rootr$ of $\statevar_{\rootr,\stateq}$ for which $\vectormu(\transvar_{\rootr,\delta})$ is non-zero, and set the corresponding entry $\vectormu^*(\transvar_{\rootr,\delta})$ in vector $\vectormu^*$ to 1 (Once again, we know that one such variable is non-zero since otherwise vector $\vectormu$ would not satisfy equation \ref{transvarConsequent}). Finally we set entry $\vectormu^*(\mainvar_{\rootr,\symbola})$  of vector $\vectormu^*$ to 1, where $\symbola = \symbolfunction{\delta}$, and note that the corresponding entry $\vectormu(\mainvar_{\rootr,\symbola})$ is necessarily non-zero otherwise $\vectormu$ would not satisfy equation \ref{transvarSymbol}. For all other variables in $\ancvars_\rootr$ we set the corresponding entry of $\vectormu^*$ to 0. 
    
    We note that this assignment $\vectormu^*_{\ancvars_\nodeu}$ satisfies inequalities (\ref{eq:2}) through (\ref{transvarSymbol}) (note that inequalities (\ref{transvarAntecedent}) are satisfied trivially since $\rootr$ has no parent) and the support of $\vectormu^*$ is contained in the support of $\vectormu$ by construction.
    
    \textbf{Inductive step: } Let $\nodeu$ be an arbitrary node of $\treeT$ with parent $\nodeu'$. Since $\vectormu^*$ is 0/1 up to node $\nodeu'$ by IH, Corollary \ref{uniquevarvalued1} ensures that there are unique variables $\statevar_{\nodeu',\stateq'}$, $\transvar_{\nodeu', \delta'}$ and $\mainvar_{\nodeu',\symbola'}$ such that $\vectormu^*(\transvar_{\nodeu', \delta'}) = \vectormu^*(\statevar_{\nodeu', \stateq'}) = \vectormu^*(\mainvar_{\nodeu', \symbola'}) = 1$. By IH, the corresponding entries $\vectormu(\statevar_{\nodeu', \stateq'})$, $\vectormu(\transvar_{\nodeu', \delta'})$ and $\vectormu(\mainvar_{\nodeu', \symbola'})$ are non-zero and so $\transvar_{\nodeu', \delta'}$ is an outgoing transition variable to some non-zero state variable $\statevar_{\nodeu, \stateq}$ in $\vectormu$ (otherwise inequalities \ref{transvarAntecedent} would not be satisfied). We set entry $\vectormu^*(\statevar_{\nodeu, \stateq})$ to 1. Next, we select arbitrarily an incoming transition variable $\transvar_{\nodeu,\delta} \in \transvarset_\nodeu$ of $\statevar_{\nodeu,\stateq}$ for which $\vectormu(\transvar_{\nodeu,\delta})$ is non-zero, and set the corresponding entry $\vectormu^*(\transvar_{\nodeu,\delta})$ in vector $\vectormu^*$ to 1 (at least one such variable is non-zero since otherwise vector $\vectormu$ would not satisfy equation \ref{transvarConsequent}). Finally we set entry $\vectormu^*(\mainvar_{\nodeu,\symbola})$  of vector $\vectormu^*$ to 1, where $\symbola = \symbolfunction{\delta}$, and note that the corresponding entry $\vectormu(\mainvar_{\nodeu,\symbola})$ is necessarily non-zero otherwise $\vectormu$ would not satisfy equation \ref{transvarSymbol}. For all other variables in $\ancvars_\nodeu$ we set the corresponding entry of $\vectormu^*$ to 0. 
    
    We note that this assignment $\vectormu^*_{\ancvars_\nodeu}$ satisfies inequalities (\ref{eq:2}) through (\ref{transvarSymbol}) and the support of $\vectormu^*$ is contained in the support of $\vectormu$ by construction. Since $\vectormu^*$ is integral and $\vectormu$ is non integral, the support of $\vectormu^*$ is strictly contained in the support of $\vectormu$ by point \ref{SprimeinS} of Observation \ref{SupportAndAlpha}.
\end{proof} 

The next lemma states that any non-integral optimal solution can be decomposed as a convex combination of a an integral solution and of a non-integral solution of smaller support. 

\begin{lemma}\label{lemma:TwoSubsolutions}
    Let $\automaton$ be a  $\treeT$-shaped tree automaton and let $\vectormu$ be a solution to $\mip_\automaton$. If $\vectormu$ is non-integral, then there is a 0/1 solution $\vectormu^*$, and a distinct feasible solution $\vectormu'$ such that
    $\vectormu$ is a convex combination of $\vectormu^*$ and $\vectormu'$. 
\end{lemma}
\begin{proof}
    Using Lemma \ref{lemma:SubsolutionIntegral} we obtain integral solution $\vectormu^*$, with support set $S^*$ strictly contained in the support set $S$ of $\vectormu$. Let  $\alpha$ be the smallest non-zero entry in $\vectormu$ chosen among the support set of $\vectormu^*$,  $\alpha = \min\{\vectormu(i)\;:\; i\in S^*\}$. 
    By point $\ref{alphalessthan1}$ of Observation \ref{SupportAndAlpha}, we know that $0< \alpha < 1$.
    
    Let $\alpha \vectormu^*$ be the vector with value $\alpha$ at every coordinate, $i \in S^*$ and zero in all other entries.  Also, let 
    $$\vectormu' =  \frac{\vectormu - \alpha\vectormu^*}{1-\alpha}$$
    
    We will show that $\vectormu'$ satisfies $\mip_\automaton$.
    
    We first observe that by Corollary \ref{EntriesLessThan1} and by our choice of alpha, that vector $\vectormu - \alpha\vectormu^*$ satisfies equalities (\ref{eq:2}). 
    
    Further, vector $\vectormu - \alpha\vectormu^*$ also satisfies (\ref{nonFinalState}) since the support of $\vectormu^*$ is contained in the support of $\vectormu$. This means that for every entry in $\vectormu$ that is $0$, corresponding entry in $\vectormu^*$ is also $0$. Therefore, the values of state variables taking part in equation (\ref{nonFinalState}) remain $0$ after $\alpha\vectormu^*$ is subtracted from $\vectormu$.
    
    Furthermore, we show that vector $\vectormu - \alpha\vectormu^*$ satisfies equalities (\ref{transvarConsequent})-(\ref{transvarSymbol}). To see this, note that Corollary \ref{uniquevarvalued1} guarantees that at most one state variable in $\alpha\vectormu^*$ takes value $\alpha$ for each node $\nodeu \in \nodes{\treeT}$. Moreover, by Lemma \ref{vectortoterm} vector $\vectormu^* = \tracehat$ for some trace $\trace$ of some term in $\languagemap(\automaton)$. Therefore by definition \ref{definition:VectorOfATrace} of vector $\tracehat$, for each $\nodeu \in \nodes{\treeT}$, state variable $\statevar_{\nodeu, \stateq}$ is equal to $1$ if and only if exactly one summand in  $\sum_{\delta}\transvar_{\nodeu,\delta}$ of equality (\ref{transvarConsequent}) is equal to $1$. The same is true for state variables and summands in $\sum_{\delta}\transvar_{\parent{\nodeu},\delta}$ of equalities (\ref{transvarAntecedent}) and for main variables and summands $\sum_{\delta}\transvar_{\nodeu,\delta}$ of equalities (\ref{transvarSymbol}). Therefore, equalities (\ref{transvarConsequent} - \ref{transvarSymbol}) hold for vector $\vectormu-\alpha\vectormu^*$ since value $\alpha$ (or $0$) is subtracted exactly once from both sides of each equality in (\ref{transvarConsequent} - \ref{transvarSymbol}) when difference $\vectormu - \alpha\vectormu^*$ is taken. 
    
    By a similar argumentation, sum $\sum \statevar_{\rootr,\stateq}$, of equality (\ref{finalState}) equals $1-\alpha$ and therefore does not hold under assignment $\vectormu-\alpha\vectormu^*$. However, equality (\ref{finalState}) holds precisely upon scaling $\vectormu-\alpha\vectormu^*$ by $1/(1-\alpha)$. Since equations (\ref{nonFinalState}) - (\ref{transvarSymbol}) are homogeneous, they too remain satisfied upon non-negative scaling (see \cite{BoydVandenberghe2004} chapter 2). 
    
    Lastly, it remains to show that inequalities (\ref{eq:2}) remain satisfied upon scaling $\vectormu-\alpha\vectormu^*$ by $1/(1-\alpha)$. By homogeneity all variables of $\vectormu-\alpha\vectormu^*$ remain greater than or equal to $0$ upon scaling. To see that no variable becomes greater than $1$, note that for each $\nodeu$ the sum of all transition variables in $\statevarset_\nodeu$ valued under $\vectormu$ is equal to $1$ by Lemma \ref{SumEntriesLessThan1}. Meanwhile, by Corollary \ref{uniquevarvalued1} exactly one variable in $\statevarset_\nodeu$ is valued $1$ under $\vectormu^*$. Therefore, by our choice of $\alpha$ no variable in $\statevarset_\nodeu$ can be larger than $1-\alpha$ under $\vectormu - \alpha\vectormu^*$. The same holds for variables in $\transvarset_\nodeu$ and $\mainvarset_\nodeu$ by analogous argumentation. Hence all variables of $\vectormu-\alpha\vectormu^*$ remain less than or equal to $1$ upon scaling by $1/(1-\alpha)$
    and inequalities (\ref{eq:2}) also hold for $\vectormu'$.
    
    Since all equalities of $\mip_\automaton$ are satisfied by $\vectormu'$ and 
    
    $$\vectormu = \alpha\cdot \vectormu^* + \left(1-\alpha\right)\cdot\vectormu',$$
    we conclude that $\vectormu$ is a convex combination of solutions $\vectormu^*$ and $\vectormu'$. 
\end{proof}

\begin{lemma}\label{lemma:NoNonintegralSolution}
    Every vertex of the polytope $\polytopeP(\mip_{\automaton})$ is
    a $0/1$-vector.  
\end{lemma}
\begin{proof}
    The proof is by contradiction. Let $\vectormu$ be a vertex of  $\polytopeP(\mip_{\automaton})$ and assume that $\vectormu$ is non-integral. By lemma  \ref{lemma:TwoSubsolutions} $\vectormu$ can be decomposed into a convex combination of distinct feasible solutions in $\polytopeP(\mip_{\automaton})$. By Definition \ref{VerticesCannotBeDecomoposed} this is a contradiction since $\vectormu$ was assumed to be a vertex and therefore cannot be decomposed into a combination of distinct feasible solutions in $\polytopeP(\mip_{\automaton})$.
\end{proof}

\section{Witness Trees and Solution Polytopes}
In this section, we let $\problem$ be a vertex subset problem, $\dpcore$ be a DP-core solving $\problem$, $\graphG$ be a graph and $\treedec$ be a tree decomposition of $\graphG$.

\begin{definition}\label{definition:WitnessTree}
    A $\dpcore$-witness-tree for $\treedec$ is a function $\witnesstree:\nodes{\tree}\rightarrow \{0,1\}^*$ satisfying the following properties:
    \begin{enumerate}
        \item $\final(\witnesstree(\rootfunction{\tree})) = 1$, 
        \item $\witnesstree(\nodeu) \in \leafcore$ if $\nodeu$ is a leaf node,
        \item If $\nodeu$ is a node with a single child $\nodeu'$,
        \begin{enumerate}
            \item $\witnesstree(\nodeu)\in\introvertexcore(\vertexv,\witnesstree(\nodeu'))$ if vertex $\vertexv$ is introduced at node $\nodeu$,  \item $\witnesstree(\nodeu)\in \forgetvertexcore(\vertexv,\witnesstree(\nodeu'))$ if vertex $\vertexv$ is forgotten at node $\nodeu$, 
            \item $\witnesstree(\nodeu) \in \introedgecore(\vertexv,\vertexv',\witnesstree(\nodeu'))$
            if an edge $\edgee$ with 
            $\epoints{\edgee} = \{\vertexv,\vertexv'\}$ is introduced at node $\nodeu$.
        \end{enumerate}
        \item If $\nodeu$ is a node with children $\nodeu'$ and $\nodeu''$, then $\witnesstree(\nodeu)\in \joincore(\witnesstree(\nodeu'),\witnesstree(\nodeu''))$. 
    \end{enumerate}
\end{definition}

Intuitively, a $\dpcore$-witness-tree is a certificate that the tree decomposition $\treedec$ is accepted by $\dpcore$. One pertinent question is whether one can devise a way of extracting a solution for $\problem$ in $\graphG$ from a witness tree. 

\begin{definition}\label{definition:AbstractMembership}
    An {\em abstract membership function} is any function of type $\membershipfunction:\N\times \{0,1\}^* \rightarrow \{0,1\}$. We say that $\vertexv\in \N$ is a  
    $\membershipfunction$-member of $\witness$ if $\membershipfunction(\vertexv,\witness) = 1$. 
\end{definition}

Let   $\vertexnodemap[\graphG,\treedec]:\vertexset_{\graphG}\rightarrow \nodes{\tree}$ be the function that assigns to each vertex $\vertexv\in \vertexset_{\graphG}$, the child of the node where $\vertexv$ is forgotten. We denote this node by $\vertexnodemap[\graphG,\treedec](\vertexv)$. This function is well defined, since in a nice edge-introducing tree decomposition, for each vertex $v$ there is exactly one node $\nodeu$ where $v$ is forgotten. We can extract a subset of vertices from a witness tree as follows.
\[
\extractedvertexset(\graphG,\treedec,\witnesstree,\membershipfunction) = \{\vertexv \in \vertexset_{\graphG}\;:\; \membershipfunction(\vertexv,\witnesstree(\vertexnodemap[\graphG,\treedec](\vertexv))) = 1 \}.
\]

Intuitively, $\extractedvertexset(\graphG,\treedec,\witnesstree,\membershipfunction)$ is the set of all vertices $\vertexv$ of $\graphG$ that are $\mu$-members of the string assigned by $\witnesstree$ to the child of the node of $\treedec$ where $\vertexv$ is forgotten. 

\begin{definition}\label{definition:SolutionPreserving}
    We say that $\dpcore$ is {\em solution-preserving} if there is an abstract membership function $\membershipfunction$ such that for each graph $\graphG$ and each 
    nice edge-introducing tree decomposition $\treedec$ accepted by $\dpcore$, the following conditions are satisfied: 
    \begin{enumerate}
        \item\label{solutionPreservingOne} For each witness tree $\witnesstree$ of $\treedec$, the set $\extractedvertexset(\graphG,\treedec,\witnesstree,\membershipfunction)$ belongs to $\solutionset_{\problem}(\graphG)$. 
        \item\label{solutionPreservingTwo} For each solution 
        $X\in \solutionset_{\problem}(\graphG)$, 
        there is a witness tree $\witnesstree$ such that 
        $X =\extractedvertexset(\graphG,\treedec,\witnesstree,\membershipfunction)$. 
    \end{enumerate}
\end{definition}

Intuitively, the first condition implies that if a solution for $\problem$ in $\graphG$ exists, then such a solution can be retrieved by backtracking: after inductively constructing the set $\dynamization(\nodeu)$ for each $\nodeu\in \nodes{\treeT}$, and determining that $\dynamization(\rootfunction{\treeT})$ is non-empty, we may construct a witness-tree $\witnesstree$ by backtracking and then apply the membership function to the string associated with the child of each forget node to extract a solution. 
On the other hand, Condition \ref{solutionPreservingTwo} ensures that each solution in $\solutionset_{\problem}(\graphG)$ can be retrieved from some suitable 
witness-tree.

\begin{definition}\label{definition:CharacteristicTree}
    Let $\graphG$ be a graph, $\treedec=(\treeT,\bagfunction,\edgefunction)$ be a tree decomposition of $\graphG$, and $\solsetX\subseteq \vertexset_{\graphG}$. The 
    characteristic tree of $\solsetX$ is 
    the $\treeT$-shaped term $\characteristictree[\graphG,\treedec,\solsetX]$ where for each $\nodeu \in \nodes{\treeT}$,  
    $\characteristictree[\graphG,\treedec,\solsetX](\nodeu) = 1$ if there is some 
    $\vertexv \in \solsetX$ with 
    $\vertexnodemap[\graphG,\treedec](\vertexv) = 
    \nodeu$, and 
    $\characteristictree[\graphG,\treedec,\solsetX](\nodeu) = 0$, otherwise. 
\end{definition}

Given a problem $\problem$, and a graph $\graphG$, we let 
$$\allcharacteristictrees_{\problem}(\graphG,\treedec) = 
\{\characteristictree[\graphG,\treedec,\solsetX]\;:\; \solsetX\in \solutionset_{\problem}(\graphG)\}$$
be the set of characteristic trees associated with solutions of $\graphG$. The next theorem establishes a connection between DP-cores deciding a problem $\problem$ and $\treeT$-shaped tree automata accepting the set of characteristic 
trees of solutions of a graph $\graphG$. 

\begin{theorem}\label{theorem:FromDPCoresToTreeAutomata}
    Let $\problem$ be a vertex subset problem and $\dpcore$ be a solution-preserving DP-core of table-complexity $\alpha(\treewidthvalue,n)$ solving $\problem$. Let $\graphG$ be a graph and $\treedec = (\treeT,\bagfunction,\edgefunction)$ be a tree decomposition of $\graphG$ of width $\treewidthvalue$. Then, there is a
    $\treeT$-shaped tree automaton $\automaton$
    of width $\alpha(\treewidthvalue,n)$ accepting the tree language $\allcharacteristictrees_{\problem}(\graphG,\treedec)$.
\end{theorem}

The proof of Theorem \ref{theorem:FromDPCoresToTreeAutomata} is carried out in detail in the next subsection.

As a direct consequence of Corollary 
\ref{corollary:MainCorollary} and Theorem \ref{theorem:FromDPCoresToTreeAutomata} we have our main 
result establishing an upper bound on extension complexity in terms of the complexity of a DP-core.

\begin{theorem}\label{theorem:DPToExtendedFormulation}
    Let $\problem$ be a vertex subset problem and $\dpcore$ be a solution-preserving DP-core of table-complexity $\alpha(\treewidthvalue,n)$ solving $\problem$. For each $n$-vertex graph $\graphG$ of treewidth $\treewidthvalue$, the extension complexity of $\polytopeP_{\problem}(\graphG)$ is at most $O(\alpha(\treewidthvalue,n)\cdot n)$. 
\end{theorem}
\begin{proof}
    Follows directly from Corollary \ref{corollary:MainCorollary} and Theorem \ref{theorem:FromDPCoresToTreeAutomata}.
\end{proof}

\subsection{Proof of Theorem \ref{theorem:FromDPCoresToTreeAutomata}}
In this section we prove Theorem \ref{theorem:FromDPCoresToTreeAutomata}, which states that given a 
solution-preserving DP-core of table-complexity $\alpha(\treewidthvalue,n)$ solving a vertex-problem $\problem$, together with a width-$\treewidthvalue$ tree decomposition 
$\treedec = (\treeT,\bagfunction,\edgefunction)$ 
of a graph $\graphG$, one can construct a $\treeT$-shaped tree automaton $\automaton$
of width $\alpha(\treewidthvalue,n)$ accepting the set of all characteristic trees of solutions of $\problem$ in $\graphG$. This result, combined with Theorem \ref{theorem:MainTheorem} provides us with a method of defining extended formulations of complexity $O(\alpha(\treewidthvalue,n)\cdot n)$ for the solution polytope of $\problem$ in $\graphG$, that is, for the polytope $\polytopeP_{\problem}(\graphG)$. 

\begin{proof}
    Let $\graphG$ be a graph and $\treedec = (\treeT,\bagfunction,\edgefunction)$ be a 
    tree decomposition of $\graphG$.
    Let $\problem$ be a vertex problem, and $\dpcore$ be a solution-preserving DP-core of table complexity $\alpha(\treewidthvalue,n)$ solving $\problem$ with abstract membership function $\membershipfunction$. 
     We define a  $\tree$-shaped automaton $\automaton$, where for 
    each node $\nodeu\in \nodes{\treeT}$, 
    $\stateset_{\nodeu} = \{q_{u,\witness}\;:\; \witness\in \dynamization(\nodeu)\}$. 
    The set of final states is 
    $\finalstateset = \{q_{\rootfunction{\tree},\witness}\;:\; \final(\witness)=1 \}$. 
    Now, the transition relation is defined as follows: 
    \begin{enumerate}
        \item If $\nodeu$  is a leaf node, then for each string $\witness\in \leafcore$ we add the arity-0 transition
        $(0,q_{\nodeu,\witness})$ to $\Delta_{\nodeu}$. 
        \item If $\nodeu$ is a node with a unique child $\nodeu'$ and (unique) parent $\nodeu''$, then for each string $\witness'\in \dynamization(\nodeu')$, and each string $\witness\in \dynamization(\nodeu)$: 
        \begin{enumerate}
            \item if vertex $v$ is introduced at node $\nodeu$ in $\treedec$ and $\witness\in \introvertexcore(\vertexv,\witness')$, then:
            \begin{enumerate}
                \item  we add transition $(q_{\nodeu',\witness'},\membershipfunction(\vertexv'',\witness), q_{\nodeu,\witness})$ to $\Delta_{\nodeu}$, if vertex $\vertexv''$ is forgotten at $\nodeu''$, 
                \item we add transition $(q_{\nodeu',\witness'},0,q_{\nodeu,\witness})$ to $\Delta_{\nodeu}$, otherwise. 
            \end{enumerate}
            
            \item if edge $\{\vertexv,\vertexv'\}$ is introduced at node $\nodeu$ in $\treedec$, and $\witness\in \introedgecore(\{\vertexv,\vertexv'\},\witness')$,  then: 
            \begin{enumerate}
                \item  we add transition $(q_{\nodeu',\witness'},\membershipfunction(\vertexv'',\witness), q_{\nodeu,\witness})$ to $\Delta_{\nodeu}$, if vertex $\vertexv''$ is forgotten at $\nodeu''$, 
                \item we add transition $(q_{\nodeu',\witness'},0,q_{\nodeu,\witness})$ to $\Delta_{\nodeu}$, otherwise. 
            \end{enumerate}
            \item if $\vertexv$ is forgotten at node $\nodeu$ in $\treedec$ and $\witness\in \forgetvertexcore(\vertexv,\witness')$: 
            \begin{enumerate}
                \item  we add transition $(q_{\nodeu',\witness'},\membershipfunction(\vertexv'',\witness), q_{\nodeu,\witness})$ to $\Delta_{\nodeu}$, if vertex $\vertexv''$ is forgotten at $\nodeu''$, 
                \item we add transition $(q_{\nodeu',\witness'},0,q_{\nodeu,\witness})$ to $\Delta_{\nodeu}$, otherwise. 
            \end{enumerate}
        \end{enumerate}
        \item If $\nodeu$ is a node with children $\nodeu'$ and $\nodeu''$ then for each 
        $\witness'\in \dynamization(\nodeu')$, each $\witness''\in \dynamization(\nodeu'')$ and each $\witness \in \dynamization(\nodeu)$ with $\witness \in \joincore(\witness',\witness'')$, then 
        \begin{enumerate}
            \item  we add transition $(q_{\nodeu',\witness'},q_{\nodeu'',\witness''},\membershipfunction(\vertexv'',\witness),q_{\nodeu,\witness})$ to $\Delta_{\nodeu}$, if vertex $\vertexv''$ is forgotten at $\nodeu''$, 
            \item we add transition $(q_{\nodeu',\witness'},q_{\nodeu'',\witness''},0,q_{\nodeu,\witness})$ to $\Delta_{\nodeu}$, otherwise. 
        \end{enumerate}
    \end{enumerate}
    
    Before continuing, we remark that given that $\dpcore$ is solution-preserving, for any witness tree $\witnesstree$ produced by $\dpcore$, the set $\solsetX = \extractedvertexset(\graphG,\treedec,\witnesstree,\membershipfunction)$ belongs to $\solutionset_{\problem}(\graphG)$. We will refer to the characteristic tree
    $\characteristictree[\graphG,\treedec,\solsetX]$ as the \emph{projection} of witness tree $\witnesstree$. 
    Note that the projection of a witness tree $\witnesstree$ can be described explicitly as follows (function $\vertexnodemap^{-1}[\graphG,\treedec]$ below is the left inverse of $\vertexnodemap[\graphG,\treedec]$): 
    
    $$
    \projection_{\witnesstree} = 
    \begin{cases}
        \membershipfunction(\vertexnodemap^{-1}[\graphG,\treedec](\nodeu),\witnesstree(\nodeu)) & \text{if $\nodeu$ is a forget node} \\
              0 & \text{otherwise}
    \end{cases}
    $$
    
    Lastly, we will say that a term $\term$ is \emph{consistent with a witness tree} $\witnesstree$ if:
    \begin{enumerate}
        \item For each node $\nodeu$ of term $\term$ the state reached at $\nodeu$ in accepting $\term$ is $\stateq_{\nodeu,\witnesstree(\nodeu)}$ and;
        \item The symbol labeling this node $\nodeu$ of $\term$ is equal to $\projection_\witnesstree(\nodeu)$. 
    \end{enumerate}  
    
    Notice that showing that a term $\term$ is consistent with a witness tree $\witnesstree$ also shows that $\term$ is the projection of $\witnesstree$ by Condition 2 above.
    
    We now argue that a term $\term$ is accepted by automaton $\automaton$ if and only if $\term$ is consistent with some witness tree 
    $\witnesstree$ produced by DP-core $\dpcore$. We argue both directions of this bi-implication by induction. 
    
    Firstly, let $\term$ be a term accepted by $\automaton$. We that show that $\term$ is the projection of some witness tree $\witnesstree$. Note that $\term$ is $\treeT$-shaped. We proceed by induction on the depth, $d$, of the tree (from the root to the leaves) showing that for each node $\nodeu$ of term $\term$ the state reached at $\nodeu$ in accepting $\term$ is $\stateq_{\nodeu,\witnesstree(\nodeu)}$ and the symbol labeling this node $\nodeu$ of $\term$ is equal to $\projection_\witnesstree(\nodeu)$. 
    
    \begin{itemize}
        \item{\bf Basis step:}At depth $d = 0$ we have only the root node $\rootr$ which is necessarily a forget node. Let $(q_{\nodeu',\witness'},\alpha,\stateq_{\rootr,\witness})$ for $\alpha \in \{0,1\}$ be the transition taken at this node. By the definition of the transition relation of $\automaton$, $\stateq_{\rootr,\witness} \in \finalstateset$ and $\final(\witness)=1$. Therefore, by definition of a witness tree, $\witness = \witnesstree(\rootr)$ for some witness tree  $\witnesstree$. Moreover, by point 2(c) in the definition of the transition relation $\alpha =\membershipfunction(\vertexv,\witness) = \membershipfunction(\vertexnodemap^{-1}[\graphG,\treedec](\rootr),\witnesstree(\rootr)) = \projection_\witnesstree(\rootr)$, where $\vertexv$ is the vertex being forgotten at this node. So we may conclude that $\term$ is consistent with some witness tree up to this node.
        
        \item{\bf Ind. step:} Assume that all nodes of $\term$ at depth $d \leq n-1$ are consistent with some witness tree, $\witnesstree$, and consider an introduce node $\nodeu'$, with parent $\nodeu$, introducing $\vertexv$ at depth $n$. By the induction hypothesis and the definition of $\Delta_\nodeu$ we know that for some $\witness' \in \dynamization(\nodeu')$,  the transition used in accepting $\term$ at $\nodeu$ is $(\stateq_{\nodeu',\witness'},\projection_\witnesstree(\nodeu), \stateq_{\nodeu,\witnesstree(\nodeu)})$, where $\witnesstree(\nodeu) \in \introvertexcore(\vertexv,\witness')$ (if $\nodeu$ is a join node we may consider both children of $\nodeu$ at the same time and show that the induction hypothesis is satisfied with the same arguments that follow). By definition of a witness tree, for any such witness $\witness'$ there exists a witness tree $\witnesstree'$ (not necessarily equal to $\witnesstree$) for which $\witnesstree'(\nodeu') = \witness'$ and $\witnesstree'(\nodet) = \witnesstree(\nodet)$ for any node $\nodet$ at depth $d \leq n-1$. Finally by the definition of $\Delta_{\nodeu'}$ the transition taken at $\nodeu'$ is labeled $0$ which is equal to $\projection_{\witnesstree'}$ since $\nodeu'$ is not a forget vertex node.
    
        A very similar argument holds for introduce edge nodes, and join nodes. 
        For a forget node $\nodeu$ we check that the transition used at this node, $\delta$, is correctly labeled by observing that $\symbolfunction{\delta} =\membershipfunction(\vertexv,\witness) = \membershipfunction(\vertexnodemap^{-1}[\graphG,\treedec](\nodeu),\witnesstree(\nodeu)) = \projection_\witnesstree(\nodeu)$, where $\vertexv$ is the vertex being forgotten at this node and where $\witness$ and $\witnesstree$ are a witness and witness tree for which $\witnesstree(\nodeu) = \witness$.
    \end{itemize}
    
    Next consider a characteristic tree $\term = (\treeT, \lambda) = \characteristictree[\graphG,\treedec,\solsetX]$ that is the projection of some witness tree $\witnesstree$ produced by our DP-core, $\characteristictree[\graphG,\treedec,\solsetX] = \projection_\witnesstree$. We look to show that any such characteristic tree will be accepted by our above defined automaton. In this regard, we introduce the notion of a trace being \emph{valid} with respect to a witness tree $\witnesstree$ produced by our DP-core. We say that a trace $\trace$ is valid with respect to witness tree $\witnesstree$ if in addition to $\trace$ being a trace of $\term$ we require explicitly that for any node $\nodeu \in \nodes{\treeT}$, $\trace(\nodeu) = \stateq_{\nodeu,\witnesstree(\nodeu)}$.  \\
    
    In particular, we will show by induction on the height $h$ of $\treeT$ (from the leaves to the root), that for any node $\nodeu$, there exists a trace valid with respect to $\witnesstree$ up to node $\nodeu$.
    
    \noindent{\bf Basis step:} Consider a leaf node $\nodeu$ at height $0$. By the definition of $\Delta$, for any $\witness \in \leafcore$, the transition $(0,\stateq_{\nodeu,\witness})$ is present in $\Delta_\nodeu$. Since $\witnesstree(\nodeu) \in \leafcore$ by definition of a witness tree, transition $\delta = (0,\stateq_{\nodeu,\witnesstree(\nodeu)})$ is present in $\Delta_\nodeu$. Moreover, since $\symbolfunction{\delta} = 0 = \projection_\witnesstree(\nodeu)$ trace $\trace$ is indeed valid with respect to $\witnesstree$ up to this node.\\
        
    \noindent{\bf Ind. step:} Let $\nodeu$ be a node at height $\height = n$ and assume that trace $\trace$ is valid with respect to $\witnesstree$ for all nodes at height $n-1$.
    \begin{itemize}
        \item If $\nodeu$ is a leaf node, we argue in the exact same way as in the basis step.
        \item Let $\nodeu$ be an introduce vertex node with child $\nodeu'$ introducing vertex $\vertexv$. By the definition of $\Delta_\nodeu$ for $\witness' \in \dynamization(\nodeu')$ and $\witness \in \dynamization(\nodeu)$ where $\witness\in\introvertexcore(\vertexv,\witness')$, transition $(\stateq_{\nodeu',\witness'}),0,\stateq_{\nodeu,\witness})$ is present in $\Delta_\nodeu$. By the definition of set $\dynamization(\nodeu)$ and definition of a witness tree, we know that $\witnesstree(\nodeu) \in \dynamization(\nodeu)$, $\witnesstree(\nodeu') \in \dynamization(\nodeu')$ and $\witnesstree(\nodeu) \in \introvertexcore(\vertexv,\witnesstree(\nodeu'))$. Hence, transition $\delta = (\stateq_{\nodeu',\witnesstree(\nodeu')},0,\stateq_{\nodeu,\witnesstree(\nodeu)})$ is present in $\Delta_\nodeu$. As in the basis step, since $\symbolfunction{\delta} = 0 = \projection_\witnesstree(\nodeu)$ trace $\trace$ is indeed valid with respect to $\witnesstree$ up to this node.
        \item For introduce edge nodes the argument is identical by replacing vertex $\vertexv$ with edge $\{\vertexv,\vertexv'\}$ and replacing function $\introvertexcore$ with $\introedgecore$.
        \item Let $\nodeu$ be a forget vertex node with child $\nodeu'$ forgetting vertex $\vertexv$. By the definition of $\Delta_\nodeu$ for $\witness' \in \dynamization(\nodeu')$ and $\witness \in \dynamization(\nodeu)$ where $\witness\in\forgetvertexcore(\vertexv,\witness')$, transition $\membershipfunction(\vertexv,\witness)(\stateq_{\nodeu',\witness'})\rightarrow \stateq_{\nodeu,\witness}$, is present in $\Delta_\nodeu$. By the definition of set $\dynamization(\nodeu)$ and definition of witness tree, we know that $\witnesstree(\nodeu) \in \dynamization(\nodeu)$, $\witnesstree(\nodeu') \in \dynamization(\nodeu')$ and $\witnesstree(\nodeu) \in \forgetvertexcore(\vertexv,\witnesstree(\nodeu'))$. Hence, transition $\delta = \membershipfunction(\vertexv,\witnesstree(\nodeu))(\stateq_{\nodeu',\witnesstree(\nodeu')})\rightarrow \stateq_{\nodeu,\witnesstree(\nodeu)}$ is present in $\Delta_\nodeu$. As in the basis step, since $\symbolfunction{\delta} = \membershipfunction(\vertexv,\witnesstree(\nodeu)) = \projection_\witnesstree(\nodeu)$ trace $\trace$ is indeed valid with respect to $\witnesstree$ up to this node.
        \item Let $\nodeu$ be an join node with children $\nodeu'$ and $\nodeu''$. By the definition of $\Delta_\nodeu$ for $\witness' \in \dynamization(\nodeu')$, $\witness'' \in \dynamization(\nodeu'')$ and $\witness \in \dynamization(\nodeu)$ where $\witness\in\joincore(\witness',\witness'')$, transition $0(\stateq_{\nodeu',\witness'}, \stateq_{\witness',\witness''})\rightarrow \stateq_{\nodeu,\witness}$ is present in $\Delta_\nodeu$. By the definition of set $\dynamization(\nodeu)$ and definition of witness tree  we know that $\witnesstree(\nodeu) \in \dynamization(\nodeu)$, $\witnesstree(\nodeu') \in \dynamization(\nodeu')$, $\witnesstree(\nodeu'') \in \dynamization(\nodeu'')$ and $\witnesstree(\nodeu) \in \joincore(\witnesstree(\nodeu'),\witnesstree(\nodeu''))$. Hence, transition $\delta = 0(\stateq_{\nodeu',\witnesstree(\nodeu')},\stateq_{\nodeu'',\witnesstree(\nodeu'')})\rightarrow \stateq_{\nodeu,\witnesstree(\nodeu)}$ is present in $\Delta_\nodeu$. As in the basis step, since $\symbolfunction{\delta} = 0 = \projection_\witnesstree(\nodeu)$ trace $\trace$ is indeed valid with respect to $\witnesstree$ up to this node.
    \end{itemize}
\end{proof}

\subsection{Extension to Edge Subset Problems}
Although our main theorems will be stated in terms of vertex subset problems, analogous results also hold for edge subset problems. 

\begin{definition}[Edge Subset Problem]\label{definition:TupleSetProblem}
    An \emph{edge subset problem} is a subset $\problem \subset \graphs \times 
    \finitesubsets{\N}$ satisfying the following conditions: 
    \begin{enumerate}
        \item $\solsetX\subseteq \edgeset_{\graphG}$, and 
        \item for each graph $\graphH$ isomorphic to $\graphG$, and each isomorphism $\phi=(\phi_1,\phi_2)$ from $\graphG$ to $\graphH$, $(\graphH,\phi_2(\solsetX))\in \problem$. 
    \end{enumerate}
\end{definition}

Given an edge subset problem $\problem$, and instance $(\graphG,\solsetX)$ of $\problem$, the set of $\solutionset_{\problem}(\graphG)$ of all solutions of $\graphG$ is defined as in Equation \ref{equation:SolutionSet}. 

Our techniques of previous sections generalize straightforwardly to edge-subset problems. In this context, the goal is to find a set of edges satisfying a given condition. The only modification we need to do in the proofs is to use an edge-localization function instead of the vertex localization function used in the context of vertex problems. The edge-localization function is defined as the function
\[
\edgenodemap[\graphG,\treedec]:\edgeset_{\graphG}\rightarrow \nodes{\tree}\]
that assigns to each edge $\edgee\in \edgeset_{\graphG}$ the node of $\tree$ where edge $\edgee$ is introduced.
Note that $\edgenodemap[\graphG,\treedec](\edgee) = \edgefunction(\edgee)$ for each edge $\edgee\in \edgeset_{\graphG}$. Intuitively, this allows us to identify in a characteristic tree what are the coordinates that correspond to edges of a graph. 

The second modification is that instead of using the set $\extractedvertexset(\graphG,\treedec,\witnesstree,\membershipfunction)$ as done for vertex problems, we use the set  
\[
\extractededgeset(\graphG,\treedec,\witnesstree,\membershipfunction) = \{\edgee\;:\; \membershipfunction(\edgee,\witnesstree(\edgenodemap[\graphG,\treedec](\edgee))) = 1 \}.\]
This is the set of all edges that are members of witnesses assigned to intro-edge nodes of $\treedec$. Here, membership is specified by the membership function $\membershipfunction$.

\begin{theorem}\label{theorem:DPToExtendedFormulationEdge}
    Let $\problem$ be an edge subset problem and $\dpcore$ be a solution-preserving DP-core of table-complexity $\alpha(\treewidthvalue,n)$ solving $\problem$. For each $n$-vertex graph $\graphG$ of treewidth $\treewidthvalue$, the extension complexity of $\polytopeP_{\problem}(\graphG)$ is at most $O(\alpha(\treewidthvalue,n)\cdot n)$. 
\end{theorem}

\subsection{Tuple Problems}
Vertex and edge problems may be generalized straightforwardly to the setting where solutions are tuples containing sets of vertices, sets of edges, or both. We define tuple subset problems below for the case where solutions consist of $d_1$ subsets of vertices and $d_2$ subsets of edges:

\begin{definition}[Tuple Subset Problem]\label{definition:EdgeSetProblem}
    A \emph{Tuple subset problem} is a subset $\problem \subset \graphs \times 
    {\finitesubsets{\N}}^{d_1 + d_2}$ satisfying the following conditions: 
    \begin{enumerate}
        \item $\solsetX=(\solsetX_1,\dots,\solsetX_{d_1},\solsetX_{d_1+1}, \ldots, \solsetX_{d_1 + d_2})$ where  $\solsetX_i \subseteq \vertexset_{\graphG}$ if $i \leq d_1$ and  $\solsetX_i\subseteq \edgeset_{\graphG}$ otherwise,
        \item for each graph $\graphH$ isomorphic to $\graphG$, and each isomorphism $\phi=(\phi_1,\phi_2)$ from $\graphG$ to $\graphH$, $(\graphH,\phi(\solsetX))\in \problem$ where $\phi(\solsetX) = ((\phi_1\solsetX_1),\dots,\phi_1(\solsetX_{d_1}),\phi_2(\solsetX_{d_1+1}), \ldots, \phi_2(\solsetX_{d_1 + d_2}))$. 
    \end{enumerate}
\end{definition}

Given a tuple subset problem $\problem$, and instance $(\graphG,\solsetX)$ of $\problem$, the set of $\solutionset_{\problem}(\graphG)$ of all solutions of $\graphG$ is defined as in Equation \ref{equation:SolutionSet}, where $\solsetX$ denotes a tuple instead of a set and $\problem\subset \graphs \times 
{\finitesubsets{\N}}^{d_1 + d_2}$.

Our proofs and techniques further generalize to tuple subset problems. In this context, the goal is to find tuples comprising of sets of vertices, sets of edges, or both, satisfying a given condition. In the case where solutions consist of $d$-tuples comprising $d_1$ sets of vertices and $d_2$ sets of edges, we need only redefine abstract membership function $\membershipfunction$ to consist of $d = d_1 + d_2$ functions, $\membershipfunction = (\vectormu_1, \ldots, \vectormu_d)$. For tuple problem with solutions of the form $\solsetX=(\solsetX_1,\dots,\solsetX_{d}$ function $\membershipfunction_i$ for $i \leq d_1$ indicates the membership of vertices in set $\solsetX_i$ and function $\membershipfunction_j$ for $d_1<j \leq d_2$ indicates the membership of a edges in set $\solsetX_j$. We may then extract solutions from a witness tree with the tuple, 
\[
\extractedvertexset(\graphG,\treedec,\witnesstree,\membershipfunction) = (\solsetX_1, \ldots,\solsetX_d)
\]
where,
\begin{itemize}
    \item $\nodeu \in \solsetX_i \mbox{ iff } \membershipfunction_i(\nodeu,\witnesstree(\vertexnodemap[\graphG,\treedec](\nodeu))) = 1,$
    \item $\edgee \in \solsetX_j \mbox{ iff }\membershipfunction_j(\edgee,\witnesstree(\edgenodemap[\graphG,\treedec](\edgee))) = 1$
\end{itemize}

This is the tuple $(\solsetX_1, \ldots,\solsetX_d)$ where $\solsetX_i$, for $i \leq d_1$, is the set of all vertices $\vertexv$ of $\graphG$ that are $\membershipfunction_i$-members of the string assigned by $\witnesstree$ to the child of the node of $\treedec$ where $\vertexv$ is forgotten. Moreover, $\solsetX_j$, for $d_1 < j \leq d_2$, is the set of all edges that are members of witnesses assigned to intro-edge nodes of $\treedec$ according to membership function $\membershipfunction_j$.

\begin{theorem}\label{theorem:DPToExtendedFormulationTuple}
    Let $\problem$ be an tuple subset problem and $\dpcore$ be a solution-preserving DP-core of table-complexity $\alpha(\treewidthvalue,n)$ solving $\problem$. For each $n$-vertex graph $\graphG$ of treewidth $\treewidthvalue$, the extension complexity of $\polytopeP_{\problem}(\graphG)$ is at most $O(\alpha(\treewidthvalue,n)\cdot n)$. 
\end{theorem}

\section{Upper Bounds for Extended Formulations}\label{section:UpperBounds}

A wide variety of vertex subset problems can be solved by dynamic programming algorithms operating on tree decompositions. In many cases, such algorithms can be formalized using solution-preserving dynamic programming cores. In this section we show how upper bounds on the table-complexity of classical dynamic programming algorithms parameterized by treewidth can be translated into parameterized upper bounds on the extension complexity of polytopes associated with several well studied combinatorial problems on graphs. 
It turns out that for several of these problems \citep{DBLP:journals/eatcs/LokshtanovMS11} the upper bounds are asymptotically optimal under the exponential time hypothesis (ETH) \citep{DBLP:journals/jcss/ImpagliazzoP01} the strong exponential time hypothesis (SETH) \citep{DBLP:journals/jcss/ImpagliazzoP01}, or related conjectures \citep{calabro2009complexity,DBLP:conf/innovations/CarmosinoGIMPS16}. 

\begin{lemma}[\cite{BlueBook} Theorem 7.9 - 7.10]\label{BestKnownAlgs}
    Given an $n$-vertex graph of treewidth $k$,

    \begin{enumerate}
        \item {\sc IndependentSet$_\ell$} can be solved in time $2^{k}n^{\bigO{1}}$\label{BestKnownAlgsIS}
        \item {\sc DominatingSet$_{\ell}$} can be solved in time $3^{k}n^{\bigO{1}}$\label{BestKnownAlgsDS}
        \item {\sc HamiltonianCycle} can be solved in time $2^{k\log k}n^{\bigO{1}}$\label{BestKnownAlgsHam}
        \item {\sc Cut$_\ell$} can be solved in time $2^{k}n^{\bigO{1}}$\label{BestKnownAlgsCut}
        \item {\sc $d$-Coloring} can be solved in time $d^{k}n^{\bigO{1}}$\label{BestKnownAlgsCol}
    \end{enumerate}
\end{lemma}

All problems in Lemma \ref{BestKnownAlgs} above admit DP-cores with table complexity matching the stated runtimes (see appendix \ref{DP-Cores}). 

The following Lemma outlines the known lower bounds under the exponential time hypothesis for various problems parameterized by treewidth.

\begin{lemma}\label{ETHLowerBounds}
    Given an $n$-vertex graph of treewidth $k$,
    
    \begin{enumerate}
        \item {\sc IndependentSet$_\ell$} cannot be solved in time $2^{\smallo{k}}n^{\bigO{1}}$ \label{ETHLowerBoundsIS} (\cite{DBLP:journals/eatcs/LokshtanovMS11} Thm 6.2)
        \item {\sc DominatingSet$_{\ell}$} cannot be solved in time $2^{\smallo{k}}n^{\bigO{1}}$\label{ETHLowerBoundsDS} (\cite{DBLP:journals/eatcs/LokshtanovMS11} Thm 6.2)
        \item {\sc HamiltonianCycle} cannot be solved in time $2^{\smallo{k}}n^{\bigO{1}}$\label{ETHLowerBoundsHam} (\cite{BlueBook} Thm 14.6)
        \item {\sc Cut$_\ell$} cannot be solved in time $2^{\smallo{k}}n^{\bigO{1}}$\label{ETHLowerBoundsCut} (\cite{Garey1976Simplified} Thm 1.2)
        \item {\sc $d$-Coloring} cannot be solved in time $d^{\smallo{k}}n^{\bigO{1}}$\label{ETHLowerBoundsCol} (\cite{DBLP:journals/eatcs/LokshtanovMS11} Thm 6.3)
    \end{enumerate}

     unless ETH fails.
\end{lemma}

\subsection{Vertex Problems}
We say that a vertex subset problem $\problem$ has extension complexity $f(k,n)$ on $n$-vertex graphs of treewidth at most $k$ if for each $n$-vertex graph $\graphG$ of treewidth at most $k$, the polytope 
$\polytopeP_{\problem}(\graphG)$ has extension complexity at most $f(k,n)$. On the other hand, we say that $\problem$ has no extended formulation with $g(k,n)$ inequalities if no polytope $\polytopeP'$ that is an extended formulation of $\polytopeP_{\problem}(\graphG)$ can be defined with $g(k,n)$ inequalities or less. 

\begin{theorem}\label{theorem:IndependentSet}
    The problem {\sc IndependentSet$_\ell$} has 
    extension complexity $2^{k}n^{O(1)}$ on $n$-vertex graphs of treewidth at most $k$. Under ETH, this problem has no extended formulation with $2^{o(k)}n^{O(1)}$ inequalities. 
\end{theorem}
\begin{proof}
    The extension complexity follows from Theorem \ref{theorem:DPToExtendedFormulation} and point \ref{BestKnownAlgsIS} of Lemma \ref{BestKnownAlgs}. The lower bound under ETH follows from point \ref{ETHLowerBoundsIS} of Lemma \ref{ETHLowerBounds}.
\end{proof}

Let {\sc DominatingSet$_{\ell}$} be the problem consisting of all pairs of the form $(G,X)$ where $G$ is a graph and $X$ is a dominating set in $G$ of size at most $\ell$. That is, each vertex of $G$ is either in $X$ or connected to some vertex in $X$. 

\begin{theorem}\label{theorem:DominatingSet}
    The problem {\sc DominatingSet$_\ell$} has extension complexity $3^{k}n^{O(1)}$ on $n$-vertex graphs of treewidth at most $\treewidthvalue$. Under ETH, this problem has no extended formulation with $2^{o(k)}n^{O(1)}$ inequalities. 
\end{theorem}
\begin{proof}
    The extension complexity follows from Theorem \ref{theorem:DPToExtendedFormulation} and point \ref{BestKnownAlgsDS} of Lemma \ref{BestKnownAlgs}. The lower bound under ETH follows from point \ref{ETHLowerBoundsDS} of Lemma \ref{ETHLowerBounds}.
\end{proof}

Interestingly, the two results above show that our general conversion from solution-preserving DP-cores to extended formulations is optimal in general when it comes to the dependency on the table-complexity of the DP-algorithm. 

\begin{theorem}\label{OptimalConversion}
    Under ETH, the complexity $O(\alpha(k,n)\cdot n)$ in Theorem \ref{theorem:DPToExtendedFormulation} cannot be improved to $\alpha(k,n)^{o(1)}\cdot n^{O(1)}$. 
\end{theorem}
\begin{proof}
    Follows from Theorem \ref{theorem:IndependentSet} and the well known fact that linear programs are polynomial time solvable in the number of input inequalities.
\end{proof}

\subsection{Edge Problems}
Let {\sc HamiltonianCycle} be the edge subset problem consisting of all pairs of the form $(G,X)$ where $G$ is a graph and $X$ is the subset of edges of some Hamiltonian cycle. This problem can be solved by a solution-preserving DP-core of table-complexity $2^{O(k\log k)}\cdot n^{O(1)}$ on $n$-vertex graphs of treewidth at most $k$. 

\begin{theorem}\label{theorem:HamiltonianCycle}
    The problem {\sc HamiltonianCycle} has extension complexity  $2^{O(k\log k)}\cdot n^{O(1)}$ 
    on $n$-vertex graphs of treewidth at most $\treewidthvalue$. Under ETH, this problem has no extended formulation with $2^{o(k)}n^{O(1)}$ inequalities.  
\end{theorem}
\begin{proof}
    The extension complexity follows from Theorem \ref{theorem:DPToExtendedFormulationEdge} and point \ref{BestKnownAlgsHam} of Lemma \ref{BestKnownAlgs}. The lower bound under ETH follows from point \ref{ETHLowerBoundsHam} of Lemma \ref{ETHLowerBounds}.
\end{proof}

Interestingly, {\sc HamiltonianCycle} has a (non-solution-preserving) DP-core whose table-complexity, $2^{O(k)}\cdot n^{O(1)}$, matches the conditional lower bound. Although we can still extract a linear system of inequalities of size $2^{O(k)}\cdot n^{O(1)}$ from the specification of this DP-core, not all Hamiltonian cycles in the corresponding graph are guaranteed to occur as extremal solutions of this system of inequalities. 

Let $\graphG$ be a graph and $(A,B)$ be a partition of the vertex set of $\graphG$. The cut-set of this partition is the set of all edges with one endpoint in $A$ and another endpoint in $B$. We let {\sc Cut$_{\ell}$} be the edge problem consisting of all pairs of the form $(G,X)$ where $G$ is a graph and $X$ is a cut-set in $G$ of size at least $\ell$. 

\begin{theorem}
\label{theorem:Cut}
The problem {\sc Cut$_\ell$} has extension complexity $2^{k}n^{O(1)}$. Under ETH, this problem has no extended formulation with $2^{o(k)}n^{O(1)}$ inequalities. 
\end{theorem}
\begin{proof}
    The extension complexity follows from Theorem \ref{theorem:DPToExtendedFormulationEdge} and point \ref{BestKnownAlgsCut} of Lemma \ref{BestKnownAlgs}. The lower bound under ETH follows from point \ref{ETHLowerBoundsCut} of Lemma \ref{ETHLowerBounds}.
\end{proof}

\subsection{Tuple Problems}
Recall that vertex and edge problems may be generalized straightforwardly to the setting where solutions are tuples containing sets of vertices, sets of edges, or both. 
For example, if a solution is a $d$-tuple 
$X=(X_1,X_2,\dots,X_d)$ of subsets of vertices, then the vector corresponding to $X$ is the $0/1$-matrix $\hat{X}$ with $d$ rows and $|\vertexset_{\graphG}|$ columns where for each $i\in [d]$ and $v\in \vertexset_{\graphG}$, $\hat{X}_{iv} = 1$ if and only if vertex $v$ belongs to the set $X_i$. The solution polytope of such a problem is the convex hull of all such matrices.
 One prominent example of 
tuple of vertex-subsets problem is the {\sc $d$-Coloring}
problem, consisting of all pairs of the form $(G,(X_1,\dots,X_d))$ where $G$ is a graph and $(X_1,\dots,X_d)$ is a partition of the vertex set of $G$ such that each $X_i$ is an independent set in $\graphG$. 

\begin{theorem}\label{theorem:Coloring}
    The problem {\sc $d$-Coloring} has extension complexity $d^{k}n^{O(1)}$ on $n$-vertex graphs of treewidth at most $k$. Under ETH, this problem has no extended formulation with $d^{o(k)}n^{O(1)}$ inequalities. 
\end{theorem}
\begin{proof}
    The extension complexity follows from Theorem \ref{theorem:DPToExtendedFormulationTuple} and point \ref{BestKnownAlgsCol} of Lemma \ref{BestKnownAlgs}. The lower bound under ETH follows from point \ref{ETHLowerBoundsCol} of Lemma \ref{ETHLowerBounds}.
\end{proof}

\section{Lower Bounds from XC}\label{section:LowerBoundsFromEF}
Our main theorem states that solution-preserving dynamic programming algorithms with small table-complexity can be translated into extended formulations with a small number of inequalities. Interestingly, this translation allows us to obtain unconditional lower bounds on the table-complexity of solution-preserving DP-cores for a given problem from unconditional lower bounds obtained in the context of the theory of extended formulations. For instance, it has been shown in \citep{Goos2018} that there is a suitable sequence of 
graphs $\{G_n\}_{n\in \N}$ and suitable $\ell(n) \in O(n)$ such that for each $n\in \N$, the graph $G_n$ has $n$ vertices and the solution polytope $P_G$ of 
the problem 
{\sc IndependentSet$_{\ell(n)}$} has extension complexity $2^{\Omega(n/\log n)}$ on $G_n$. 

\begin{theorem}[\cite{Goos2018} Theorem 1.1] \label{ISextLB}
    There is an (explicit) family of $n$-node graphs $G$ with $\texttt{xc}(P_G) \leq 2^{\Omega(n/ \log n)}$.
\end{theorem}

This implies the following unconditional parameterized lower bound on the table-complexity of solution-preserving DP-cores solving the independent set problem. 

\begin{theorem}
    The problem {\sc IndependentSet$_{\ell}$} has no solution-preserving DP-core of table-complexity $2^{o(k/log k)}\cdot n^{O(1)}$ on $n$-vertex graphs of treewidth at most $k$. 
\end{theorem}
\begin{proof}
    Follows from Theorem \ref{theorem:DPToExtendedFormulation} and Theorem \ref{ISextLB}.
\end{proof}

\section{Conclusion}
In this work, we introduced a general framework to obtain parameterized upper bounds on the extension complexity of the solution polytopes of a wide variety of combinatorial problems in terms of the table-complexity of solution-preserving dynamic programming algorithms. In many cases, the extension complexity obtained using our method is the best one can get (under ETH). 

Nevertheless, for some combinatorial problems, such as {\sc HamiltonianCycle} and other connectivity problems, the fastest known dynamic programming algorithm parameterized by treewidth \citep{bodlaender2015deterministic} is not solution preserving. In these cases, our result still yields a polytope associated with the convex-hull of all solutions that can be retrieved by backtracking from the DP tables constructed by the algorithm. The crucial difference is that this polytope does not depend only on the graph, but also on the input tree decomposition and on the particularities of the DP-core. 

Another point that is worth mentioning is that our upper bounds on the extension complexity of polytopes in terms of the table-complexity of DP-cores does not depend on the actual computational complexity of constructing the tables. Only the table sizes matter. In this sense, it is not possible to generalize our results to arbitrary non-solution preserving DP-cores without making further assumptions. The reason is that any vertex subset problem has a trivial (non-solution preserving) DP-core of table-complexity $1$. This DP-core non-deterministically guesses a solution, and when processing each node, keeps only the partial solution corresponding to the restriction of the guessed solution to vertices belonging to bags up to that node. 

\bibliographystyle{elsarticle-harv} 
\bibliography{Bibliography}

\appendix
\section{Dynamic Programming Cores}\label{DP-Cores}
\noindent{\bf A DP-core for Dominating Set.}
Let {\sc DominatingSet$_{\ell}$} be the problem consisting of all pairs of the form $(G,X)$ where $G$ is a graph and $X$ is a dominating set in $G$ of size at most $\ell$. That is, each vertex of $G$ is either in $X$ or connected to some vertex in $X$. The DP-core that follows is modeled against the standard dynamic programming algorithm solving {\sc DominatingSet$_{\ell}$}, a detailed exposition of which can be found in \cite{BlueBook} (Chapter 7). We remark again here that this standard DP algorithm admits a runtime of $4^{O(k)}\cdot n^{O(1)}$ on $n$-vertex graphs of treewidth at most $k$. However, for the results of our work only the table complexity, $3^{O(k)}\cdot n^{O(1)}$, of this algorithm is relevant. 

We define below a DP-core $\dpcore$ for vertex subset problem {\sc DominatingSet$_{\ell}$}. We specify the operation of each component of the core on strings that encode 3 tuples of the form $(S_1,S_2,c)$, where $S_1,S_2\subseteq \N$ and $c\in \N$. The DP-core  components are defined as follows. 

\begin{itemize}
    \item $\leafcore = \{(\emptyset,\emptyset,0)\}$. 
    \item $\introvertexcore(v,(S_1,S_2,c)) = \{(S_1\cup \{v\},S_2\cup \{v\},c),(S_1,S_2,c+1)\}$.
    \item $\introedgecore(v,v',(S_1,S_2,c)) =
    \left\{\begin{array}{l}
    \{(S_1,S_2,c)\} \mbox{ if $v,v' \in S_1 \lor v,v' \notin S_1$} \\ 
    \{(S_1,S_2\cup \{v\},c)\} \mbox{ if $v \notin S_1$} \\
    \{(S_1,S_2\cup \{v'\},c)\} \mbox{ otherwise.}
    \end{array}\right.$ 
    
    \item $\forgetvertexcore(v,(S_1,S_2,c)) = 
    \left\{\begin{array}{l}
    \{\emptyset\} \mbox{ if $v \notin S_2$} \\ 
    \{(S_1\setminus \{v\},S_2\setminus \{v\},c)\} \mbox{ otherwise}
    \end{array}\right.$ 
\end{itemize}

\begin{itemize}
    \item
    {\small 
    ${\joincore((S_1,S_2,c),(S'_1,S'_2,c'))=}
    \left\{
    \begin{array}{l}
    \emptyset \mbox{ if $S_1\neq S'_1$} \\ 
    \{(S_1,S_2 \cup S'_2,c+c' - |S_1| )\} \mbox{ otw.} 
    \end{array}\right.
    $
    }
    \item $\final((S_1,S_2,c)) = 1$ if and only if $c\leq \ell$. 
\end{itemize}

\noindent{\bf A DP-core for Cut.}
Let $\graphG$ be a graph and $(A,B)$ be a partition of the vertex set of $\graphG$. The cut-set of this partition is the set of all edges with one endpoint in $A$ and another endpoint in $B$. We let {\sc Cut$_{\ell}$} be the edge problem consisting of all pairs of the form $(G,X)$ where $G$ is a graph and $X$ is a cut-set in $G$ of size at least $\ell$.
The DP-core that follows is modeled against the standard dynamic programming algorithm solving {\sc Cut$_{\ell}$}, resulting in a DP-core of table complexity of $2^{O(k)}\cdot n^{O(1)}$ on $n$-vertex graphs of treewidth at most $k$. A detailed exposition of the algorithm can be found in \cite{Bodlaender1994}.

We define below a DP-core $\dpcore$ for edge subset problem {\sc Cut$_{\ell}$}. We specify the operation of each component of the core on strings that encode 3 tuples of the form $(R,S,c)$ where $R \subseteq \N \times \N$, $S\subseteq \N$ and $c\in \N$. The DP-core  components are defined as follows. 

\begin{itemize}
    \item $\leafcore = \{(\emptyset,\emptyset,0)\}$. 
    \item $\introvertexcore(v,(R,S,c)) = \{(R,S,c),(R,S\cup \{v\},c)\}$.
    
    \item $\introedgecore(v,v',(R,S,c)) =
    \left\{\begin{array}{l}
    \{(R\cup \{v,v'\},S,c+1)\} \mbox{ if $v \in S \oplus v' \in S$} \\ 
    \{(R,S,c)\} \mbox{ otherwise.} 
    \end{array}\right.$ 
    
    \item $\forgetvertexcore(v,(R,S,c)) = \{(\{u,u'\}\in R \;|\; u\neq v \wedge u'\neq v\},S\backslash \{v\},c)\}$. 
\end{itemize}

\begin{itemize}
    \item
    {\small 
    ${\joincore((R_1,S_1,c_1),(R_2,S_2,c_2))=}
    \left\{
    \begin{array}{l}
    \emptyset \mbox{ if $S_1\neq S_2$} \\ 
    \{(R_1 \cup R_2,S_1,c_1+c_2 )\} \mbox{ otw.} 
    \end{array}\right.
    $
    }
    \item $\final((R,S,c)) = 1$ if and only if $c\geq \ell$. 
\end{itemize}

\noindent{\bf A DP-core for Hamiltonian Cycle.}
Let {\sc HamiltonianCycle} be the edge subset problem consisting of all pairs of the form $(G,X)$ where $G$ is a graph and $X$ is the subset of edges of some Hamiltonian cycle. This problem can be solved by a solution-preserving DP-core of table-complexity $2^{O(k\log k)}\cdot n^{O(1)}$ on $n$-vertex graphs of treewidth at most $k$. The DP-core is modeled against the standard solution-preserving dynamic programming algorithm solving {\sc HamiltonianCycle} (see \cite{marx2020treewidth} for a detailed exposition). 
We define below a DP-core $\dpcore$ for edge subset problem {\sc HamiltonianCycle}. We specify the operation of each component of the core on strings that encode 3 tuples of the form $(S_0,S_1,S_2)$ where $S_1 \subseteq \N \times \N$, $S_0, S_2 \subseteq \N$ and $c\in \N$. For a tuple $S = (S_0,S_1,S_2)$ we define indicator function $S[\cdot] : S_0 \cup S_1 \cup S_2 \rightarrow \{0,1,2\}$ where $S[v] = 0$ if $v$ is in $S_0$, $S[v] = 1$ if $v$ takes part in some tuple in $S_1$, and $S[v] = 2$ if $v$ is in $S_2$. The DP-core  components are defined as follows. 

\begin{itemize}
    \item $\leafcore = \{(\emptyset,\emptyset,\emptyset)\}$. 
    \item $\introvertexcore(v,(S_0,S_1,S_2)) = \{(S_0\cup \{v\},S_1,S_2)\}$.
    
    \item $\introedgecore(v,v',(R,S,c)) =
    \left\{\begin{array}{l}
    \{(S_0\setminus \{v,v'\},S_1\cup (v, v'),S_2)\} \mbox{ if $v, v' \in S_0$} \\ 
    \{(S_0\setminus v,[S_1 \setminus \{v',u\}] \cup \{v,u\},S_2)\} \mbox{ if $v \in S_0, \{v',u\} \in S_1$ (Vice versa)} \\
    \{(S_0\setminus v,[S_1 \setminus \{v',u\}] \cup \{v,u\},S_2)\} \mbox{ if $\{v,w\},\{v',u\} \in S_1, v\neq v' \neq u \neq w$} \\ 
    \{(S_0,S_1,S_2)\} \mbox{ otherwise.}
    \end{array}\right.$ 
    
    \item $\forgetvertexcore(v,(S_0,S_1,S_2)) = 
    \left\{\begin{array}{l}
    \{(S_0,S_1,S_2\setminus v)\} \mbox{ if $v \in S_2$} \\ 
    \emptyset \mbox{ otherwise.}
    \end{array}\right.$ 
\end{itemize}

\begin{itemize}
    \item
    {\small 
    ${\joincore((S_0,S_1,S_2),(S'_0,S'_1,S'_2))=}
    \left\{
    \begin{array}{l}
    \{(R_0, R_1, R_2)\} \mbox{ if for all $v$, $S[v] + S'[v] \leq 2$ and $S_1 \cup S'_1$ is acyclic} \\ 
    \emptyset \mbox{ otw.} 
    \end{array}\right.
    $
    }
    
    Where,
    \begin{itemize}
        \item $v \in R_0 \mbox{ iff } S[v] + S'[v] = 0$
        \item $\{v,u\} \in R_1 \mbox{ iff, } $ are endpoints of some path in $S_1 \cup S_2$ 
        \item $v \in R_2 \mbox{ iff } S[v] + S'[v] = 2$
    \end{itemize}
    \item $\final((S_0,S_1,S_2)) = 1$. 
\end{itemize}

\noindent{\bf A DP-core for d-Coloring.}
Let {\sc $d$-Coloring} be the
problem consisting of all pairs of the form $(G,(X_1,\dots,X_d))$ where $G$ is a graph and $(X_1,\dots,X_d)$ is a partition of the vertex set of $G$ such that each $X_i$ is an independent set in $\graphG$. The DP-core that follows is modeled against the standard dynamic programming algorithm solving {\sc $d$-Coloring} (see \cite{marx2020treewidth} for a detailed exposition) resulting in a DP-core of table-complexity $d^{O(k)}\cdot n^{O(1)}$ on $n$-vertex graphs of treewidth at most $k$.

We specify the operation of each component of the core on strings that encode $d$ tuples of the form $(S_1, \ldots, S_d)$ where, $S_i\subseteq \N$. The DP-core  components are defined as follows. 

\begin{itemize}
    \item $\leafcore = \{(\emptyset_1, \ldots, \emptyset_d)\}$. 
    \item $\introvertexcore(v,(S_1, \ldots, S_d)) = \{(S_1,\dots,S_i \cup \{v\},\dots,S_d)\;:\;
    i \in [d]\}$.
    
    \item $\introedgecore(v,v',(S_1, \ldots, S_d)) =
    \left\{\begin{array}{l}
    \{\emptyset\} \mbox{ if $v \in S_i \land v' \in S_i$ } \\ 
    \{(S_1, \ldots, S_d)\} \mbox{ otherwise.} 
    \end{array}\right.$ 
    
    \item $\forgetvertexcore(v,(S_1, \ldots, S_d)) = \{(S_1\setminus \{v\}, \ldots, S_d\setminus \{v\})\}$. 
\end{itemize}

\begin{itemize}
    \item
    {\small 
    ${\joincore((S_1, \ldots, S_d),(S'_1, \ldots, S'_d))=}
    \left\{
    \begin{array}{l}
    \emptyset \mbox{ if $S_i\neq S'_i$ for any $i \in [d]$} \\ 
    \{(S_1, \ldots, S_d)\} \mbox{ otw.} 
    \end{array}\right.
    $
    }
    \item $\final((S_1, \ldots, S_d)) = 1$. 
\end{itemize}

\end{document}